\documentclass[prc,twocolumn]{revtex4-1}

\usepackage{enumerate}

\usepackage{booktabs}
\usepackage{color}
\usepackage{graphicx}

\usepackage{amsmath}
\usepackage{amssymb}
\usepackage{times}
\usepackage{multirow}
\usepackage{makecell}
\def\bvec#1{\mbox{\boldmath $#1$}}

\newcommand{\what}[1]{\widehat{#1}}

\newcommand{\del}[2]{\frac{\partial #1}{\partial #2}}

\newcommand{\bra}{\langle}
\newcommand{\ket}{\rangle}

\newcommand{\idot}{\!\cdot\!}

\usepackage{ascmac}
\usepackage{fancybox}

\newcommand{\beq}{\begin{equation}}
\newcommand{\eeq}{\end{equation}}
\newcommand{\bea}{\begin{eqnarray}}
\newcommand{\eea}{\end{eqnarray}}

\def\fun#1#2{\lower3.6pt\vbox{\baselineskip0pt\lineskip.9pt
 \ialign{$\mathsurround=0pt#1\hfil##\hfil$\crcr#2\crcr\sim\crcr}}}

\begin{document}

\title{
  A Unified Theoretical Framework for HFB Resonant States:\\
  Integration of the Complex-Scaled Jost Function and Autonne-Takagi Normalization
}

\author{Kazuhito Mizuyama$^{1,2}$}
\email{corresponding author: mizukazu147@gmail.com}

\affiliation{
  \textsuperscript{1}
  Institute of Research and Development, Duy Tan University,
  Da Nang 550000, Vietnam
  \\
  \textsuperscript{2}
  Faculty of Natural Sciences,  Duy Tan University, Da Nang 550000, Vietnam
}

\date{\today}

\begin{abstract}
  We develop a theoretical framework to describe quasiparticle resonance states within the Hartree-Fock-Bogoliubov (HFB) theory by integrating the complex-scaled Jost
  function method with the Autonne-Takagi factorization. The HFB completeness relation is derived from the analytical properties of the Green's function using contour
  integration in the complex energy plane, where the complex scaling method (CSM) is shown to be essential for explicitly separating resonance pole contributions from the
  continuum background. To uniquely define and normalize the resonant wave functions (Gamow states), the Autonne-Takagi factorization is applied to the rank-1 residue
  matrix of the flux-adjusted S-matrix at the pole energy. This scheme determines the absolute scale and phase of the eigenfunctions without relying on artificial
  adjustments or phenomenological basis sets. Numerical analysis confirms that physical observables and the defined wave functions remain invariant under the rotation of
  the complex scaling angle $\theta$. Furthermore, the T-matrix residues calculated via the Mittag-Leffler expansion are shown to be in exact numerical agreement with
  those obtained from the microscopic integrals of the Takagi-normalized Gamow states. Our analysis of the scattering profiles reveals that hole-type quasiparticle
  resonances can be understood as a manifestation of the Fano process originating from the interference between the discrete poles and the background continuum. The
  proposed normalization scheme provides a foundation for evaluating the collectivity of various excitation modes in open quantum many-body systems.
\end{abstract}

\maketitle

\section{Introduction}

The unified description of nuclear structure and reactions remains one of the most challenging goals in nuclear physics. The Hartree-Fock-Bogoliubov (HFB) theory
\cite{Ring1980, Bulgac1980, Dobaczewski1984} has been highly successful as a self-consistent mean-field framework that incorporates pairing correlations, which are
essential for describing nuclei far from the stability line. In such weakly bound systems, the quasiparticle Hamiltonian naturally defines a spectrum that includes not
only bound states but also a continuous spectrum and resonances. A proper treatment of this continuum is indispensable, as the coupling between bound quasiparticle
configurations and the continuum significantly impacts the ground-state properties and excitation modes of drip-line nuclei \cite{Belyaev1987, Matsuo2001}.

Despite its success, a rigorous treatment of resonance states (Gamow states) within the HFB framework is non-trivial. The primary difficulty arises from the boundary
conditions: resonant wave functions satisfy outgoing-wave conditions that lead to exponential divergence at large distances, making the standard Hermitian normalization
and the formulation of a discrete completeness relation problematic.

To address resonances, several powerful methods have been developed. The Complex Scaling Method (CSM) \cite{ABC1, ABC2, Simon1973, Moiseyev1998, Aoyama2006} has been
widely used to isolate resonances as discrete eigenvalues of a non-Hermitian Hamiltonian. Traditionally, the CSM is implemented using basis expansion techniques (e.g.,
the harmonic oscillator basis). While efficient for calculating complex energies, the basis expansion can sometimes obscure the direct connection between the obtained
eigenstates and the analytic structure of the S-matrix defined by scattering boundary conditions. On the other hand, the Gamow Shell Model (GSM) \cite{Michel2002,Michel2009}
employs the Berggren basis \cite{Berggren1968}, which explicitly includes resonance poles in the completeness relation. However, for coupled-channel HFB
systems involving both particle and hole components, rigorously deriving such a completeness relation from the first principles of the Green's function—without assuming
a specific basis set a priori—remains a fundamental task.

In our previous work \cite{Mizuyama2019}, we developed the Jost-HFB method, which allows for the accurate location of S-matrix poles as zeros of the Jost function
determinant. This approach respects the proper boundary conditions of the HFB equation in coordinate space. However, two major problems remained: the rigorous
definition of the resonant wave function at the pole and the formulation of a completeness relation that explicitly separates resonance contributions.

In this paper, we establish a complete and self-consistent theoretical framework to resolve these issues. We achieve this by integrating the Jost-HFB method with the
CSM and the Autonne-Takagi factorization \cite{Autonne1915, Takagi1925}.
The key novelty of this work lies in the following points.
First, we provide a rigorous derivation of the HFB completeness relation by performing a contour integration of the HFB Green's function in the complex energy plane. We
demonstrate that the CSM is not merely a numerical tool but a theoretical necessity to rotate the integration contour and unmask the resonance poles from the continuum
background.
Second, we propose a mathematically unique normalization scheme for resonant wave functions. By utilizing the fact that the "flux-adjusted" S-matrix in the HFB
framework is a complex symmetric matrix \cite{Mizuyama2024}, and noting that its residue matrix at the pole is of rank-1, we apply the Autonne-Takagi factorization
(also known as symmetric singular value decomposition). This allows us to uniquely determine the absolute scale and phase of the Gamow states.
Finally, we perform a comprehensive numerical verification. We demonstrate that the T-matrix residues calculated via the Mittag-Leffler expansion \cite{Rakityansky,Mizuyama2025}
—which depends only on the analyticity of the S-matrix—are in exact agreement with the microscopic integrals of our Takagi-normalized wave functions.

The paper is organized as follows. In Sec. II, we describe the theoretical framework, including the complex-scaled HFB equation, the Jost function, the derivation of
the completeness relation, and the normalization via Takagi factorization. In Sec. III, we present the numerical results for various quasiparticle resonances, including
a discussion on the Fano-type interference and the numerical consistency of the residues. Finally, a summary and outlook are given in Sec. IV.

\section{Theory}
As given in Ref.\cite{Mizuyama2019}, the HFB equation in coordinate space is expressed as a
simultaneous second-order differential equation in the 2x2 matrix form as
\begin{eqnarray}
  \bvec{H}
  \phi_{lj}(r;E)
  =
  \frac{\hbar^2}{2m}
  \bvec{\mathcal{K}}^2(E)
  \phi_{lj}(r;E)
  \label{hfbeq}
\end{eqnarray}
using the $2\times 2$ matrix Hamiltonian defined by
\begin{eqnarray}
  \bvec{H}
  &\equiv&
  \left[
    \bvec{H}^{(0)}
    +
    \bvec{\sigma}_z
    \bvec{\mathcal{U}}
    \right]
\end{eqnarray}
with
\begin{eqnarray}
  \bvec{H}^{(0)}
  &\equiv&
  -
  \frac{\hbar^2}{2m}
  \left(\del{^2}{r^2}-\frac{l(l+1)}{r^2}\right)
  \bvec{1}
  \\
  \bvec{\mathcal{U}}
  &\equiv&
  \begin{pmatrix}
    U_{lj}(r) & \Delta(r) \\
    \Delta(r) & -U_{lj}(r)
  \end{pmatrix},
\end{eqnarray}
\begin{eqnarray}
  \bvec{\mathcal{K}}(E)
  =
  \begin{pmatrix}
    k_1(E) & 0 \\
    0 & k_2(E)
  \end{pmatrix},
\end{eqnarray}
\begin{eqnarray}
  \bvec{1}
  =
  \begin{pmatrix}
    1 & 0 \\
    0 & 1
  \end{pmatrix}
  \text{ and }
  \bvec{\sigma}_z
  =
  \begin{pmatrix}
    1 & 0 \\
    0 & -1
  \end{pmatrix},
  \label{defmat}
\end{eqnarray}
where $U_{lj}(r)$ and $\Delta(r)$ are the mean-field potential and pairing potential, respectively,
and $k_1(E)$ and $k_2(E)$ are the momentum which are defined using the complex quasiparticle
energy $E$ and
chemical potential $\lambda$, as $k_1(E)\equiv\sqrt{\frac{2m}{\hbar^2}(\lambda+E)}$
and $k_2(E)\equiv\sqrt{\frac{2m}{\hbar^2}(\lambda-E)}$.
$\phi_{lj}$ is a two-component spinor with an upper component and a lower component given as
$\phi_{lj}^{\mathsf{T}}=(\phi_{lj,1},\phi_{lj,2})$. 

\subsection{Complex scaled HFB equation}
The complex scaling method is a technique in which both the coordinate $r$ and the momentum
$k$ are extended to complex numbers. Due to the consistency of the Fourier transform and
the requirement of canonical conjugacy, the coordinate axis in the complex plane is rotated
as $r\to re^{i\theta }$, while the momentum axis is rotated in the opposite direction
as $k\rightarrow ke^{-i\theta }$. This ensures that the product $kr$ remains invariant,
so that plane‑wave factors like $e^{ikr}$ are preserved, while resonant states can be
isolated as discrete eigenvalues of the non-Hermitian Hamiltonian.

Within the framework of the Jost-HFB method, the complex scaling for $r$ is $r\to r e^{i\theta}$
as before, while for the momentum, momentum matrix $\bvec{\mathcal{K}}$ is scaled as
$\bvec{\mathcal{K}}(E)\to \bvec{\mathcal{K}}_{\theta}(E) e^{-i\theta}$.
The HFB equation with complex scaling applied is thus given by:
\begin{eqnarray}
  \bvec{H}_\theta
  \phi_{lj}^{\theta}(r;E)
  =
  \frac{\hbar^2}{2m}
  \bvec{\mathcal{K}}_{\theta}^2(E)
  \phi_{lj}^{\theta}(r;E)
  \label{hfbeq-csm}
\end{eqnarray}
where
\begin{eqnarray}
  \bvec{H}_{\theta}
  &\equiv&
  \left[
    \bvec{H}^{(0)}
    +
    \bvec{\sigma}_z
    \bvec{\mathcal{V}}_{\theta}
    \right]
  \label{hamiltonian-csm}
\end{eqnarray}
and
\begin{eqnarray}
  \bvec{\mathcal{V}}_{\theta}(r)
  &\equiv&
  e^{2i\theta}
  \bvec{\mathcal{U}}(r e^{i\theta})
  \\
  \bvec{\mathcal{K}}_{\theta}(E)
  &\equiv &
  \begin{pmatrix}
    k_1^{\theta}(E) & 0 \\
    0 & k_2^{\theta}(E)
  \end{pmatrix}
  =
  \bvec{\mathcal{K}}(E)e^{i\theta}
  \label{Kmat-csm}
\end{eqnarray}
with
\begin{eqnarray}
  k_1^{\theta}(E)
  &\equiv&
  \sqrt{\frac{2m}{\hbar^2}(\lambda+E)e^{2i\theta}}
  =k_1(E)e^{i\theta},
  \label{defk1-csm}
  \\
  k_2^{\theta}(E)
  &\equiv&
  \sqrt{\frac{2m}{\hbar^2}(\lambda-E)e^{2i\theta}}
  =
  k_2(E)e^{i\theta}.
  \label{defk2-csm}
\end{eqnarray}

Due to complex scaling, the Hamiltonian becomes non-Hermitian and the eigenstates
no longer form an orthonormal basis. Consequently, a dual basis is necessary to
consistently define the eigenexpansion and inner product.

The dual basis in the complex scaled HFB equation can be defined as solutions which
satisfy the following equation. 
\begin{eqnarray}
  \bvec{H}^*_\theta
  \widetilde{\phi}_{lj}^{\theta}(r;E)
  =
  \frac{\hbar^2}{2m}
  \widetilde{\bvec{\mathcal{K}}}_{\theta}^2(E)
  \widetilde{\phi}_{lj}^{\theta}(r;E)
  \label{hfbeq-csm-dual}
\end{eqnarray}
with a new momentum matrix $\widetilde{\bvec{\mathcal{K}}}_{\theta}$ defined by 
\begin{eqnarray}
  \widetilde{\bvec{\mathcal{K}}}_{\theta}(E)
  &\equiv &
  \begin{pmatrix}
    \widetilde{k}_1^{\theta}(E) & 0 \\
    0 & \widetilde{k}_2^{\theta}(E)
  \end{pmatrix}
  =
  \bvec{\mathcal{K}}(E)e^{-i\theta}
\end{eqnarray}
where
\begin{eqnarray}
  \widetilde{k}_1^{\theta}(E)
  &\equiv&
  \sqrt{\frac{2m}{\hbar^2}(\lambda+E)e^{-2i\theta}}
  =k_1(E)e^{-i\theta},
  \label{defk1-csm-dual}
  \\
  \widetilde{k}_2^{\theta}(E)
  &\equiv&
  \sqrt{\frac{2m}{\hbar^2}(\lambda-E)e^{-2i\theta}}
  =
  k_2(E)e^{-i\theta}.
  \label{defk2-csm-dual}
\end{eqnarray}
Since there are two regular solutions and two irregular solutions for a system of
second-order differential equations given in 2x2 matrix form,
Eq.(\ref{hfbeq-csm}) has two regular solutions ($\phi_{lj}^{(r1)\theta}$ and $\phi_{lj}^{(r2)\theta}$) and
two irregular solutions ($\phi_{lj}^{(\pm 1)\theta}$ and $\phi_{lj}^{(\pm 2)\theta}$).
Eq.(\ref{hfbeq-csm-dual}) also has two regular solutions ($\tilde{\phi}_{lj}^{(r1)\theta}$ and
$\tilde{\phi}_{lj}^{(r2)\theta}$) and two irregular solutions ($\tilde{\phi}_{lj}^{(\pm 1)\theta}$
and $\tilde{\phi}_{lj}^{(\pm 2)\theta}$).

By using two solutions for regular and irregular solutions, we can define
the $2\times 2$ matrix solutions as
\begin{eqnarray}
  \bvec{\Phi}^{(r)\theta}_{lj}
  &\equiv&
  \begin{pmatrix}
    \phi_{lj}^{(r1)\theta} &
    \phi_{lj}^{(r2)\theta}
  \end{pmatrix}
  \\
  \bvec{\Phi}^{(\pm)\theta}_{lj}
  &\equiv&
  \begin{pmatrix}
    \phi_{lj}^{(\pm 1)\theta} &
    \phi_{lj}^{(\pm 2)\theta}
  \end{pmatrix}
\end{eqnarray}
and
\begin{eqnarray}
  \widetilde{\bvec{\Phi}}^{(r)\theta}_{lj}
  &\equiv&
  \begin{pmatrix}
    \tilde{\phi}_{lj}^{(r1)\theta} &
    \tilde{\phi}_{lj}^{(r2)\theta}
  \end{pmatrix}
  \\
  \widetilde{\bvec{\Phi}}^{(\pm)\theta}_{lj}
  &\equiv&
  \begin{pmatrix}
    \tilde{\phi}_{lj}^{(\pm 1)\theta} &
    \tilde{\phi}_{lj}^{(\pm 2)\theta}
  \end{pmatrix}.
\end{eqnarray}
These solution matrices satisfy
\begin{eqnarray}
  \bvec{H}_\theta
  \bvec{\Phi}_{lj}^{\theta}
  &=&
  \frac{\hbar^2}{2m}
  \bvec{\mathcal{K}}_\theta^2(E)
  \bvec{\Phi}_{lj}^{\theta}
  \label{hfbeq-mat}
\end{eqnarray}
and
\begin{eqnarray}
  \bvec{H}_\theta^*
  \widetilde{\bvec{\Phi}}_{lj}^{\theta}
  &=&
  \frac{\hbar^2}{2m}
  \widetilde{\bvec{\mathcal{K}}}_\theta^2(E)
  \widetilde{\bvec{\Phi}}_{lj}^{\theta}
  \label{hfbeq-mat-dual}
\end{eqnarray}

Regular and irregular solutions for the original basis (Eq.(\ref{hfbeq-mat})) can be obtained
by solving with the following boundary conditions:
\begin{eqnarray}
  \lim_{r\to 0}
  \bvec{\Phi}^{(r)\theta}_{lj}
  \sim
  \bvec{\chi}_{l}^{(r)\theta}
  =
  \begin{pmatrix}
    F_l(k_1^\theta(E)r)  & 0 \\
    0 & F_l(k_2^\theta(E)r)
  \end{pmatrix}
  \label{asymPhir}
\end{eqnarray}
and
\begin{eqnarray}
  \lim_{r\to \infty}
  \bvec{\Phi}^{(\pm)\theta}_{lj}
  \sim
  \bvec{\chi}_{l}^{(\pm)\theta}
  =
  \begin{pmatrix}
    O_l^{(\pm)}(k_1^\theta(E)r)  & 0 \\
    0 & O_l^{(\pm)}(k_2^\theta(E)r)
  \end{pmatrix}.
  \label{asymPhi}
\end{eqnarray}
Regular and irregular solutions for the dual basis (Eq.(\ref{hfbeq-mat})) can be obtained
by solving with the following boundary conditions:
\begin{eqnarray}
  \lim_{r\to 0}
  \widetilde{\bvec{\Phi}}^{(r)\theta}_{lj}
  \sim
  \widetilde{\bvec{\chi}}_{l}^{(r)\theta}
  =
  \begin{pmatrix}
    F_l(\tilde{k}_1^\theta(E)r)  & 0 \\
    0 & F_l(\tilde{k}_2^\theta(E)r)
  \end{pmatrix}
  \label{asymPhir-dual}
\end{eqnarray}
and
\begin{eqnarray}
  \lim_{r\to \infty}
  \widetilde{\bvec{\Phi}}^{(\pm)\theta}_{lj}
  \sim
  \widetilde{\bvec{\chi}}_{l}^{(\pm)\theta}
  =
  \begin{pmatrix}
    O_l^{(\pm)}(\tilde{k}_1^\theta(E)r)  & 0 \\
    0 & O_l^{(\pm)}(\tilde{k}_2^\theta(E)r)
  \end{pmatrix},
  \label{asymPhi-dual}
\end{eqnarray}
where $F_l$ and $O_l^{(\pm)}$ are  defined by using the spherical Bessel and Hankel
functions as $F_l(kr)\equiv rj_l(kr)$ and $O_l^{(\pm)}(kr)\equiv rh_l^{(\pm)}(kr)$, and
$\bvec{\chi}_{l}^\theta$ and $\widetilde{\bvec{\chi}}_{l}^\theta$ satisfy the following equations:
\begin{eqnarray}
  \bvec{H}^{(0)}
  \bvec{\chi}_{l}^\theta(r;E)
  =
  \frac{\hbar^2}{2m}
  \bvec{\mathcal{K}}_{\theta}^2(E)
  \bvec{\chi}_{l}^\theta(r;E)
  \label{free-csm}
\end{eqnarray}
and
\begin{eqnarray}
  \bvec{H}^{(0)}
  \widetilde{\bvec{\chi}}_{l}^\theta(r;E)
  =
  \frac{\hbar^2}{2m}
  \widetilde{\bvec{\mathcal{K}}}_{\theta}^2(E)
  \widetilde{\bvec{\chi}}_{l}^\theta(r;E)
  \label{free-csm-dual},
\end{eqnarray}
respectively.

\subsection{Wronskian and Jost function}
Since $\bvec{\mathcal{V}}_\theta$ is symmetric
($\bvec{\mathcal{V}}_\theta^{\mathsf{T}}=\bvec{\mathcal{V}}_\theta$)
in the Hamiltonian given by Eq.(\ref{hamiltonian-csm}),
we have the following identity for any independent solution matrix
$\bvec{\Phi}_{lj}^{(1)\theta}(E)$ and and $\bvec{\Phi}_{lj}^{(2)\theta}(E)$,
\begin{eqnarray}
  \int_a^b dr
  \bvec{\Phi}_{lj}^{(1)\theta\mathsf{T}}(E)
  \left[
    \bvec{H}_{\theta}^{\mathsf{T}}
    \bvec{\sigma}_z
    -
    \bvec{\sigma}_z
    \bvec{H}_{\theta}
    \right]
  \bvec{\Phi}_{lj}^{(2)\theta}(E)
  =0
  \label{identity}
\end{eqnarray}
where the differential operator included in $\bvec{H}_\theta^{\mathsf{T}}$ shall
be applied to the left-hand solution matrix. Applying the Green's theorem
to the l.h.s of Eq.(\ref{identity}), we obtain
\begin{eqnarray}
  &&
  \left.\bvec{\mathcal{W}}\left(\bvec{\Phi}_{lj}^{(1)}(E),\bvec{\Phi}_{lj}^{(2)}(E)\right)\right|_{r=b}
  \nonumber\\
  &&\hspace{20pt}
  -
  \left.\bvec{\mathcal{W}}\left(\bvec{\Phi}_{lj}^{(1)}(E),\bvec{\Phi}_{lj}^{(2)}(E)\right)\right|_{r=a}
  =0
  \label{identity2}.
\end{eqnarray}
where the Wronskian defined by
\begin{eqnarray}
  &&
  \bvec{\mathcal{W}}\left(\bvec{\Phi}_{lj}^{(1)\theta}(E),\bvec{\Phi}_{lj}^{(2)\theta}(E)\right)
  \nonumber\\
  &&
  \equiv
  \frac{\hbar^2}{2m}
  \left[
    \bvec{\Phi}_{lj}^{(1)\theta\mathsf{T}}(E)
    \bvec{\sigma}_z
    \del{\bvec{\Phi}_{lj}^{(2)\theta}(E)}{r}
    \right.
    \nonumber\\
    &&\hspace{40pt}
    \left.
    -
    \del{\bvec{\Phi}_{lj}^{(1)\theta\mathsf{T}}(E)}{r}
    \bvec{\sigma}_z
    \bvec{\Phi}_{lj}^{(2)\theta}(E)
    \right].
  \label{wron-rp}
\end{eqnarray}
Since $a$ and $b$ can be taken arbitrarily as long as the solution does not diverge there,
we see that the Wronskian defined by Eq.(\ref{wron-rp}) is, as is generally well known,
independent of the coordinate $r$.

If we take $\bvec{\Phi}_{lj}^{(-)\theta}(E)$ as $\bvec{\Phi}_{lj}^{(1)\theta}(E)$
and $\bvec{\Phi}_{lj}^{(+)\theta}(E)$ as $\bvec{\Phi}_{lj}^{(2)\theta}(E)$
in $\bvec{\mathcal{W}}\left(\bvec{\Phi}_{lj}^{(1)\theta}(E),\bvec{\Phi}_{lj}^{(2)\theta}(E)\right)$,
since the value of $\bvec{\mathcal{W}}\left(\bvec{\Phi}_{lj}^{(-)\theta}(E),
\bvec{\Phi}_{lj}^{(+)\theta}(E)\right)$ is determined by the asymptotic form at infinity
for $\bvec{\Phi}_{lj}^{(-)\theta}(E)$ and $\bvec{\Phi}_{lj}^{(-)\theta}(E)$
(Eq.(\ref{asymPhi})) due to the independence on $r$, we obtain
\begin{eqnarray}
  \bvec{\mathcal{W}}\left(\bvec{\Phi}_{lj}^{(-)\theta}(E),\bvec{\Phi}_{lj}^{(+)\theta}(E)\right)
  &=&
  \bvec{\mathcal{W}}\left(\bvec{\chi}_{l}^{(-)\theta}(E),\bvec{\chi}_{l}^{(+)\theta}(E)\right)
  \nonumber\\
  &=&
  i\frac{\hbar^2}{m}
  \bvec{\sigma}_z
  \left(\bvec{\mathcal{K}}_\theta(E)\right)^{-1}
  \label{wron-rp-Phi}
\end{eqnarray}
(Note that this is not an approximation.)

The complex scaled Jost function is defined as a $2\times 2$ matrix
function which connects the complex scaled regular and irregular solution matrices as
\begin{eqnarray}
  \bvec{\Phi}_{lj}^{(r)\theta\mathsf{T}}
  &=&
  \frac{1}{2}
  \left[
    \bvec{\mathcal{J}}_{lj}^{(+)\theta}
    \bvec{\Phi}_{lj}^{(-)\theta\mathsf{T}}
    +
    \bvec{\mathcal{J}}_{lj}^{(-)\theta}
    \bvec{\Phi}_{lj}^{(+)\theta\mathsf{T}}
    \right].
  \label{defjost-csm}
\end{eqnarray}
By using Eqs.(\ref{wron-rp-Phi}) and (\ref{defjost-csm}), we can obtain
\begin{eqnarray}
  &&
  \bvec{\mathcal{J}}_{lj}^{(\pm)\theta}(E)
  \nonumber\\
  &&=
  \pm
  \frac{1}{i}
  \frac{2m}{\hbar^2}
  \bvec{\mathcal{W}}\left(\bvec{\Phi}_{lj}^{(r)\theta}(E),\bvec{\Phi}_{lj}^{(\pm)\theta}(E)\right)
  \bvec{\mathcal{K}}_\theta(E)
  \bvec{\sigma}_z
  \label{wron-jost1-csm}.
\end{eqnarray}
Also, since there is a relation
\begin{eqnarray}
  \bvec{\mathcal{W}}\left(\bvec{\Phi}_{lj}^{(r)},\bvec{\Phi}_{lj}^{(\pm)}\right)
  =
  e^{i\theta}
  \bvec{\mathcal{W}}\left(\bvec{\Phi}_{lj}^{(r)\theta},\bvec{\Phi}_{lj}^{(\pm)\theta}\right),
\end{eqnarray}
we can see that the Jost function, as a function of complex energy, is invariant under
complex scaling; that is,
\begin{eqnarray}
  \bvec{\mathcal{J}}_{lj}^{(\pm)\theta}(\bvec{\mathcal{K}}_\theta(E))
  =
  \bvec{\mathcal{J}}_{lj}^{(\pm)}(\bvec{\mathcal{K}}(E)).
  \label{invarianceJost}
\end{eqnarray}
Similarly, in the dual space, the Wronskian for the irregular solution matrix is given as:
\begin{eqnarray}
  \bvec{\mathcal{W}}\left(\widetilde{\bvec{\Phi}}_{lj}^{(-)\theta}(E),
  \widetilde{\bvec{\Phi}}_{lj}^{(+)\theta}(E)\right)
  &=&
  i\frac{\hbar^2}{m}
  \bvec{\sigma}_z
  \left(\widetilde{\bvec{\mathcal{K}}}_\theta(E)\right)^{-1}
  \label{wron-rp-Phi-dual}
\end{eqnarray}
and since the definition of the Jost function is given as 
\begin{eqnarray}
  \widetilde{\bvec{\Phi}}_{lj}^{(r)\theta\mathsf{T}}
  &=&
  \frac{1}{2}
  \left[
    \widetilde{\bvec{\mathcal{J}}}_{lj}^{(+)\theta}
    \widetilde{\bvec{\Phi}}_{lj}^{(-)\theta\mathsf{T}}
    +
    \widetilde{\bvec{\mathcal{J}}}_{lj}^{(-)\theta}
    \widetilde{\bvec{\Phi}}_{lj}^{(+)\theta\mathsf{T}}
    \right],
  \label{defjost-csm-dual}
\end{eqnarray}
we can obtain
\begin{eqnarray}
  &&
  \widetilde{\bvec{\mathcal{J}}}_{lj}^{(\pm)\theta}(E)
  \nonumber\\
  &&=
  \pm
  \frac{1}{i}
  \frac{2m}{\hbar^2}
  \bvec{\mathcal{W}}
  \left(\widetilde{\bvec{\Phi}}_{lj}^{(r)\theta}(E),\widetilde{\bvec{\Phi}}_{lj}^{(\pm)\theta}(E)\right)
  \widetilde{\bvec{\mathcal{K}}}_\theta(E)
  \bvec{\sigma}_z
  \label{wron-jost1-csm-dual}.
\end{eqnarray}

\subsection{Integral representation of the Jost function}
The integral form of the Jost function can be obtained from the following identity:
\begin{eqnarray}
  \lim_{\epsilon\to 0}
  \int_\epsilon^{\infty} dr
  \bvec{\chi}_{l}^{(\pm)\theta\mathsf{T}}(E)
  \left[
    \bvec{H}^{(0)\mathsf{T}}
    \bvec{\sigma}_z
    -
    \bvec{\sigma}_z
    \bvec{H}_{\theta}
    \right]
  \bvec{\Phi}_{lj}^{(r)\theta}(E)
  =0.
  \nonumber\\
  \label{identity3}
\end{eqnarray}
This identity can be written as
\begin{eqnarray}
  &&
  \left.\bvec{\mathcal{W}}\left(\bvec{\chi}_{l}^{(\pm)\theta}(E),
  \bvec{\Phi}_{lj}^{(r)\theta}(E)\right)\right|_{r=\infty}
  \nonumber\\
  &&
  =
  \lim_{\epsilon\to 0}
  \left[
    \left.\bvec{\mathcal{W}}\left(\bvec{\chi}_{l}^{(\pm)\theta}(E),
    \bvec{\Phi}_{lj}^{(r)\theta}(E)\right)\right|_{r=\epsilon}
    \right.
    \nonumber\\
    &&\hspace{30pt}
    \left.
    +
    \int_\epsilon^{\infty} dr
    \bvec{\chi}_{l}^{(\pm)\theta\mathsf{T}}(E)
    \bvec{\mathcal{V}}_{\theta}
    \bvec{\Phi}_{lj}^{(r)\theta}(E)
    \right]
  \label{int-eq}
\end{eqnarray}
By using Eqs.(\ref{asymPhir}), (\ref{asymPhi}) and (\ref{defjost-csm}),
we can obtain
\begin{eqnarray}
  &&
  \left.\bvec{\mathcal{W}}\left(\bvec{\chi}_{l}^{(\pm)\theta}(E),
  \bvec{\Phi}_{lj}^{(r)\theta}(E)\right)\right|_{r=\infty}
  \nonumber\\
  &&\hspace{20pt}
  =
  \mp i
  \frac{\hbar^2}{2m}
  \left(\bvec{\mathcal{K}}_{\theta}(E)\right)^{-1}
  \bvec{\sigma}_z
  \bvec{\mathcal{J}}_{lj}^{(\pm)\theta\mathsf{T}}(E)
\end{eqnarray}
and
\begin{eqnarray}
  &&
  \lim_{\epsilon\to 0}
  \left.\bvec{\mathcal{W}}\left(\bvec{\chi}_{l}^{(\pm)\theta}(E),
  \bvec{\Phi}_{lj}^{(r)\theta}(E)\right)\right|_{r=\epsilon}
  \nonumber\\
  &&\hspace{30pt}
  =
  \mp i
  \frac{\hbar^2}{2m}
  \left(\bvec{\mathcal{K}}_{\theta}(E)\right)^{-1}
  \bvec{\sigma}_z
  \label{wron-chi}
\end{eqnarray}
The reason for considering the limit as $\epsilon \to 0$ is due to the divergence of
$\chi_l^{(\pm)}$ at $r=0$. However, taking the limit after calculating the Wronskian as
in Eq.(\ref{wron-chi}) allows us to avoid this divergence. Therefore, we obtain the
integral form of the Jost function as
\begin{eqnarray}
  \bvec{\mathcal{J}}_{lj}^{(\pm)\theta\mathsf{T}}
  =
  \bvec{1}
  \mp
  \frac{1}{i}
  \frac{2m}{\hbar^2}
  \bvec{\sigma}_z
  \bvec{\mathcal{K}}_\theta
  \int_0^\infty dr
  \bvec{\chi}_{l}^{(\pm)\theta\mathsf{T}}
  \bvec{\mathcal{V}}_\theta
  \bvec{\Phi}_{lj}^{(r)\theta}
  \label{jost-int}
\end{eqnarray}
The integral form of the Jost function for the dual space is also derived as
\begin{eqnarray}
  \widetilde{\bvec{\mathcal{J}}}_{lj}^{(\pm)\theta\mathsf{T}}
  =
  \bvec{1}
  \mp
  \frac{1}{i}
  \frac{2m}{\hbar^2}
  \bvec{\sigma}_z
  \widetilde{\bvec{\mathcal{K}}}_\theta
  \int_0^\infty dr
  \widetilde{\bvec{\chi}}_{l}^{(\pm)\theta\mathsf{T}}
  \bvec{\mathcal{V}}_\theta^*
  \widetilde{\bvec{\Phi}}_{lj}^{(r)\theta}.
  \label{jost-int-dual}
\end{eqnarray}

\subsection{Symmetrization of the S-matrix (flux-adjusted S-matrix)}

The scattering wave function and the “S-matrix” are defined by multiplying
the inverse of the Jost function to the regular solution, as
\begin{eqnarray}
  &&
  \bvec{\Psi}_{lj}^{(+)\theta\mathsf{T}}(E)
  =
  \left(\bvec{\mathcal{J}}_{lj}^{(+)\theta}(E)\right)^{-1}
  \bvec{\Phi}_{lj}^{(r)\theta\mathsf{T}}(E)
  \label{defscat}
  \\
  &&=
  \frac{1}{2}
  \left[
    \bvec{\Phi}_{lj}^{(-)\theta\mathsf{T}}(E)
    +
    \bvec{\mathcal{S}}_{lj}^{\theta}(E)
    \bvec{\Phi}_{lj}^{(+)\theta\mathsf{T}}(E)
    \right]
  \label{Scatwf}
\end{eqnarray}
where the ``S-matrix'' is defined by
\begin{eqnarray}
  \bvec{\mathcal{S}}_{lj}^{\theta}(E)
  &\equiv&
  \left(\bvec{\mathcal{J}}_{lj}^{(+)\theta}(E)\right)^{-1}
  \bvec{\mathcal{J}}_{lj}^{(-)\theta}(E).
  \label{defSmat}
\end{eqnarray}
The Wronskian of two scattering wave functions
$\bvec{\mathcal{W}}\left(\bvec{\Psi}_{lj}^{(+)\theta}(E),\bvec{\Psi}_{lj}^{(+)\theta}(E)\right)$ is zero,
independent of the coordinate $r$.
Inserting Eq.(\ref{defSmat}) into this Wronskian and using Eq.(\ref{wron-rp-Phi}) yields the following
identity with respect to the ``S-matrix'' as
\begin{eqnarray}
  \bvec{\sigma}_z
  \left(\bvec{\mathcal{K}}_\theta(E)\right)^{-1}
  \bvec{\mathcal{S}}_{lj}^{\theta\mathsf{T}}(E)
  =
  \bvec{\mathcal{S}}_{lj}^{\theta}(E)
  \bvec{\sigma}_z
  \left(\bvec{\mathcal{K}}_\theta(E)\right)^{-1}
\end{eqnarray}
Although this result shows that the ‘S-matrix’ is clearly not a symmetric matrix,
analogous to the definition introduced
in Ref.~\cite{Mizuyama2024}, introducing the ``flux-adjusted S-matrix''
$\bvec{\mathcal{S}}^{\theta}_{lj,F}$ and
the ``flux-adjusted'' regular and irregular solution matrices
($\bvec{\Phi}^{(r)\theta}_{lj,F}$ and $\bvec{\Phi}^{(\pm)\theta}_{lj,F}$) as
\begin{eqnarray}
  \bvec{\Phi}^{(r)\theta\mathsf{T}}_{lj,F}
  &\equiv&
  \bvec{\mathcal{K}}_\theta^{\frac{1}{2}}\bvec{\sigma}_z^{\frac{1}{2}}
  \bvec{\Phi}^{(r)\theta\mathsf{T}}_{lj}
  \label{Phir-F}
  \\
  \bvec{\Phi}^{(\pm)\theta\mathsf{T}}_{lj,F}
  &\equiv&
  \bvec{\mathcal{K}}_\theta^{\frac{1}{2}}\bvec{\sigma}_z^{\frac{1}{2}}
  \bvec{\Phi}^{(\pm)\theta\mathsf{T}}_{lj}
  \label{Phii-F}
  \\
  \bvec{\mathcal{S}}^{\theta}_{lj,F}
  &\equiv&
  \bvec{\mathcal{K}}_\theta^{\frac{1}{2}}\bvec{\sigma}_z^{\frac{1}{2}}
  \bvec{\mathcal{S}}^{\theta}_{lj}
  \bvec{\sigma}_z^{-\frac{1}{2}}\bvec{\mathcal{K}}_\theta^{-\frac{1}{2}}
  \label{Smat-flux}
\end{eqnarray}
we can find that the ``flux-adjusted S-matrix'' which satisfies
\begin{eqnarray}
  \left(\bvec{\mathcal{S}}^{\theta}_{lj,F}\right)^{\mathsf{T}}=\bvec{\mathcal{S}}^{\theta}_{lj,F}.
  \label{Smat_prop}
\end{eqnarray}

\subsection{Negative energy solution}
The unitary matrix $(-i\sigma_y)$ keeps the Hamiltonian $\bvec{H}_{\theta}$ invariant
(i.e., $(-i\sigma_y)\bvec{H}_{\theta}(i\sigma_y)=\bvec{H}_{\theta})$ and, for the momentum matrix,
gives $(-i\sigma_y)\bvec{\mathcal{K}}_\theta(E)(i\sigma_y)=\bvec{\mathcal{K}}_\theta(-E)$.
Therefore, $(-i\sigma_y)\phi_{lj}$ can be considered to give a negative-energy solution
(a solution with inverted energy).
However, when the unitary matrix is actually applied to the regular solution,
we obtain the result as
\begin{eqnarray}
  (-i\sigma_y)
  \phi_{lj}^{(r1)\theta}(E)
  &=&
  \begin{pmatrix}
    -\phi_{2,lj}^{(r1)\theta}(E) \\
    \phi_{1,lj}^{(r1)\theta}(E)
  \end{pmatrix}
  =
  \begin{pmatrix}
    -\phi_{1,lj}^{(r2)\theta}(-E) \\
    \phi_{2,lj}^{(r2)\theta}(-E)
  \end{pmatrix}
  \nonumber\\
  &=&
  -\sigma_z
  \phi_{lj}^{(r2)\theta}(-E)
  \\
  (-i\sigma_y)
  \phi_{lj}^{(r2)\theta}(E)
  &=&
  -\sigma_z
  \phi_{lj}^{(r1)\theta}(-E)
\end{eqnarray}
by considering the boundary condition of the solution.
This can be expressed in matrix form as
\begin{eqnarray}
  (-i\sigma_y)
  \bvec{\Phi}^{(r)\theta}_{lj}(E)
  &=&
  -\sigma_z
  \bvec{\Phi}^{(r)\theta}_{lj}(-E)
  \sigma_x.
\end{eqnarray}
Therefore, this expression finally becomes as
\begin{eqnarray}
  \sigma_x
  \bvec{\Phi}^{(r)\theta}_{lj}(E)
  \sigma_x
  &=&
  \bvec{\Phi}^{(r)\theta}_{lj}(-E).
  \label{reg-neg}
\end{eqnarray}
For the irregular solution matrix, we can obtain the same formula as
\begin{eqnarray}
  \sigma_x
  \bvec{\Phi}^{(\pm)\theta}_{lj}(E)
  \sigma_x
  &=&
  \bvec{\Phi}^{(\pm)\theta}_{lj}(-E).
  \label{ireg-neg}
\end{eqnarray}
Transforming the Hamiltonian $\bvec{H}_\theta$ using $\sigma_x$ simply changes the pairing potential $\Delta(r)$ to $-\Delta(r)$.
The momentum matrix is transformed as $\sigma_x\bvec{\mathcal{K}}_\theta(E)\sigma_x=\bvec{\mathcal{K}}_\theta(-E)$,
the same as when using $-i\sigma_y$. It is known that in the HFB equations,
the solution remains unchanged even when $\Delta(r)$ is replaced by $-\Delta(r)$~\cite{}; therefore $\sigma_x$
is also a unitary matrix that provides negative-energy solutions.
\\
The energy inverted Jost function (negative energy Jost function) can be obtained as
\begin{eqnarray}
  \bvec{\mathcal{J}}^{(\pm)\theta}_{lj}(-E)
  &=&
  \sigma_x
  \bvec{\mathcal{J}}^{(\pm)\theta}_{lj}(E)
  \sigma_x
  \\
  &=&
  \begin{pmatrix}
    \bvec{\mathcal{J}}^{(\pm)\theta}_{22,lj}(E)  & \bvec{\mathcal{J}}^{(\pm)\theta}_{21,lj}(E) \\
    \bvec{\mathcal{J}}^{(\pm)\theta}_{12,lj}(E)  & \bvec{\mathcal{J}}^{(\pm)\theta}_{11,lj}(E)
  \end{pmatrix}.
  \label{jost-neg}
\end{eqnarray}
This transformation does not change the determinant of the Jost function, i.e.
\begin{eqnarray}
  \det\bvec{\mathcal{J}}^{(\pm)\theta}_{lj}(-E)
  =\det\bvec{\mathcal{J}}^{(\pm)\theta}_{lj}(E).
  \label{detjost-neg}
\end{eqnarray}
Since the pole of the S-matrix $E_n$ satisfies $\det\bvec{\mathcal{J}}^{(\pm)\theta}_{lj}(E_n)=0$,
when a pole is found at $E=E_n$, another pole also exists at $E=-E_n$.
The same argument holds in the dual space, and the same symmetric property can be derived as
\begin{eqnarray}
  \det\widetilde{\bvec{\mathcal{J}}}^{(\pm)\theta}_{lj}(-E)
  =\det\widetilde{\bvec{\mathcal{J}}}^{(\pm)\theta}_{lj}(E).
  \label{detjost-neg-dual}
\end{eqnarray}

\subsection{Riemann sheet}
Within the framework of the complex scaling HFB theory, two momenta, $k_1^\theta(E)$ and
$k_2^\theta(E)$, are defined as shown in Eqs.(\ref{defk1-csm}) and (\ref{defk2-csm}).
When an integer $n (=0$ or $1)$ is given, since $e^{2in\pi}=1$,
the complex number $(\lambda+E)e^{2i\theta}$ appearing in
the definitions of $k_1^\theta(E)$ can be expressed as:
\begin{eqnarray}
  (\lambda+E)e^{2i\theta}
  &=&
  |\lambda+E|
  e^{i{(\arg(\lambda+E)+2\theta+2n\pi)}}.
\end{eqnarray}
Therefore, $k_1^\theta(E)$ can take two values as
\begin{eqnarray}
  k_1^{\theta}(E)
  &=&
  \sqrt{\frac{2m}{\hbar^2}|\lambda+E|}e^{i{(\arg((\lambda+E)e^{2i\theta})/2+n\pi)}}
  \label{k-polar}
  \\
  &=&
  \left\{
  \begin{array}{lc}
    \sqrt{\frac{2m}{\hbar^2}|\lambda+E|}e^{i{(\arg((\lambda+E)e^{2i\theta})/2)}} & (n=0) \\
    \\
    -\sqrt{\frac{2m}{\hbar^2}|\lambda+E|}e^{i{(\arg((\lambda+E)e^{2i\theta})/2)}} & (n=1) \\
  \end{array}
  \right.
\end{eqnarray}
for a given value of $E$, depending on the value of $n$($=0$ or $1$). Such a function is called
two-valued function.

In the Re $E>0$ region, to ensure single-valuedness, we define a branch cut for $k_1^\theta(E)$ starting from
the branch point $E = -\lambda$ along the half-line with an angle $\arg(\lambda+E) = -2\theta$.
The domain for the phase is restricted to $0 < \arg((\lambda+E)e^{2i\theta}) < 2\pi$. 
Within this domain: In the $n=0$ branch, the total phase of $k_1^\theta(E)$ lies in $(0, \pi)$,
ensuring $\operatorname{Im} k_1^\theta(E) > 0$. 
In the $n=1$ branch, the total phase lies of $k_1^\theta(E)$ in $(\pi, 2\pi)$,
ensuring $\operatorname{Im} k_1^\theta(E) < 0$. 
When crossing a branch cut, the phase of $k_1^\theta(E)$ changes discontinuously from $0$
to $\pi$ (flipping from $n = 0$ to $n = 1$), whereas the phase of $k_1^\theta(E)$ changes
continuously while maintaining $\operatorname{Im} k_2^\theta(E) > 0$. 
Consequently, when a complex energy plane satisfying $\operatorname{Im} k_1^\theta(E) > 0$
and $\operatorname{Im} k_2^\theta(E) > 0$ is defined as the first Riemann sheet, a complex
energy plane satisfying $\operatorname{Im} k_1^\theta(E) < 0$ and
$\operatorname{Im} k_2^\theta(E) > 0$, which is analytically connected to the first Riemann
sheet across a branch cut, is defined as the second Riemann sheet.

In the dual space, $\tilde{k}_1^\theta(E)$ and $\tilde{k}_2^\theta(E)$ are defined by
Eqs.(\ref{defk1-csm-dual}) and (\ref{defk2-csm-dual}).
$\tilde{k}_1^\theta(E)$ can also be expressed as
\begin{eqnarray}
  \tilde{k}_1^{\theta}(E)
  &=&
  \sqrt{\frac{2m}{\hbar^2}|\lambda+E|}e^{i{(\arg((\lambda+E)e^{-2i\theta})/2+m\pi)}}
  \label{kt-polar}
  \\
  &=&
  \left\{
  \begin{array}{lc}
    \sqrt{\frac{2m}{\hbar^2}|\lambda+E|}e^{i{(\arg((\lambda+E)e^{-2i\theta})/2)}} & (m=0) \\
    \\
    -\sqrt{\frac{2m}{\hbar^2}|\lambda+E|}e^{i{(\arg((\lambda+E)e^{-2i\theta})/2)}} & (m=1) \\
  \end{array}
  \right.
\end{eqnarray}
A branch cut for $\tilde{k}_1^\theta(E)$ is defined by the half-line starting from
the branch point $E = -\lambda$ with an angle $\arg(\lambda+E) = 2\theta$.
The domain for the phase is restricted to $0 < \arg((\lambda+E)e^{-2i\theta}) < 2\pi$. 
Within this domain: In the $m=0$ branch, the total phase of $\tilde{k}_1^\theta(E)$ lies in $(0, \pi)$,
ensuring $\operatorname{Im} \tilde{k}_1^\theta(E) > 0$. 
In the $m=1$ branch, the total phase lies of $\tilde{k}_1^\theta(E)$ in $(\pi, 2\pi)$,
ensuring $\operatorname{Im} \tilde{k}_1^\theta(E) < 0$. 
When crossing a branch cut, the phase of $\tilde{k}_1^\theta(E)$ changes discontinuously from $0$
to $\pi$ (flipping from $m = 0$ to $m = 1$), whereas the phase of $\tilde{k}_1^\theta(E)$ changes
continuously while maintaining $\operatorname{Im} \tilde{k}_2^\theta(E) > 0$.
Therefore, also in the dual space, when a complex energy plane satisfying $\operatorname{Im} \tilde{k}_1^\theta(E) > 0$
and $\operatorname{Im} \tilde{k}_2^\theta(E) > 0$ is defined as the first Riemann sheet, a complex
energy plane satisfying $\operatorname{Im} \tilde{k}_1^\theta(E) < 0$ and
$\operatorname{Im} \tilde{k}_2^\theta(E) > 0$, which is analytically connected to the first Riemann
sheet across a branch cut, is defined as the second Riemann sheet.

To discuss the relationship between $\bvec{\mathcal{K}}_\theta(E)$ and $\widetilde{\bvec{\mathcal{K}}}_\theta(E)$,
we first consider the representation of $\tilde{k}_1^\theta(E^*)$ and its argument.
If we do not consider the two-valued nature of $\tilde{k}_1^\theta$, then when we replace $E$ with $E^*$
in Eq.(\ref{kt-polar}), the argument becomes $\arg((\lambda+E^*)e^{-2i\theta}) = -\arg((\lambda+E)e^{2i\theta})$. 
However, if we consider $E^*$ to be the complex conjugate of $E$ given on a specific Riemann sheet,
then $E^*$ should also belong to the same Riemann sheet (i.e., when Im $\tilde{k}_1^{\theta}(E) > 0$, then
Im $\tilde{k}_1^{\theta}(E^*) >0$, and if Im $\tilde{k}_1^{\theta}(E) < 0$, then Im $\tilde{k}_1^{\theta}(E^*) < 0$).
Therefore, there must be a relationship that 
\begin{eqnarray}
  \arg((\lambda+E^*)e^{-2i\theta}) = 2\pi - \arg((\lambda+E)e^{2i\theta}),
\end{eqnarray}
and we obtain 
\begin{eqnarray}
  \tilde{k}_1^{\theta}(E^*)
  &=&
  -\sqrt{\frac{2m}{\hbar^2}|\lambda+E|}e^{-i{(\arg((\lambda+E)e^{2i\theta})/2-m\pi)}}
  \label{kt-pola2}
  \\
  &=&
  \left\{
  \begin{array}{lc}
    -\sqrt{\frac{2m}{\hbar^2}|\lambda+E|}e^{-i{\arg((\lambda+E)e^{2i\theta})/2}} & (m=0) \\
    \\
    \sqrt{\frac{2m}{\hbar^2}|\lambda+E|}e^{-i{\arg((\lambda+E)e^{2i\theta})/2}} & (m=1) \\
  \end{array}
  \right.
\end{eqnarray}
For $\tilde{k}_2^\theta(E^*)$, since it is required that, in the region where Re $E>0$,
both Im $\tilde{k}_2^\theta(E)>0$ and Im $\tilde{k}_2^\theta(E^*)>0$ are always satisfied,
we have the following relation
\begin{eqnarray}
  \arg((\lambda-E^*)e^{-2i\theta}) = 2\pi - \arg((\lambda-E)e^{2i\theta}).
\end{eqnarray}
Therefore $k_2^\theta$ and $\tilde{k}_2^\theta$ are related as
\begin{eqnarray}
  \tilde{k}_2^\theta(E^*)=-k_2^{\theta *}(E).
\end{eqnarray}
The relationship between $\bvec{\mathcal{K}}_\theta(E)$ and $\widetilde{\bvec{\mathcal{K}}}_\theta(E)$ in Re $E > 0$ region,
derived from the above discussion can be summarised as
\begin{enumerate}
\item If $E$ is the first Riemann sheet or the second Riemann sheet defined on both $k_1^\theta$ and $\tilde{k}_1^\theta$ (i.e. $n=m=0$ or $1$),
  \begin{eqnarray}
    \widetilde{\bvec{\mathcal{K}}}_\theta(E^*)=-\bvec{\mathcal{K}}_\theta^*(E)
    \label{KKt1}
  \end{eqnarray}
\item If $E$ is different between the Riemann sheet defined by $k_1^\theta$ and
  the Riemann sheet defined by $\tilde{k}_1^\theta$ (i.e. $n \neq m$),
  \begin{eqnarray}
    \widetilde{\bvec{\mathcal{K}}}_\theta(E^*)=\bvec{\sigma}_z\bvec{\mathcal{K}}_\theta^*(E).
    \label{KKt2}
  \end{eqnarray}
\end{enumerate}
The relationship between $\bvec{\mathcal{K}}_\theta(E)$ and $\widetilde{\bvec{\mathcal{K}}}_\theta(E)$
in the negative energy region (Re $E < 0$) can be obtained by exchanging $k_1^\theta$ and $k_2^\theta$,
$\tilde{k}_1^\theta$ and $\tilde{k}_2^\theta$ in the above discussion, but since we have already shown
that there is symmetry between the positive and negative energy region in the previous section,
and since it is sufficient to consider only one of these regions when discussing scattering problems,
we omit the details here. 

\subsection{Relationship between the original and dual spaces}
\label{original_vs_dual}
Since the regular and irregular solutions obtained by solving Eqs.(\ref{hfbeq-mat})
and (\ref{hfbeq-mat-dual}) retain the symmetric properties with respect to
momentum that satisfy the boundary conditions
(Eqs.(\ref{asymPhir})-(\ref{asymPhi-dual})), 
we can obtain the relationship between the solutions (regular and irregular) in the
original space and the solutions (regular and irregular) in the dual space.
Furthermore, by applying the derived relationship to the definitions of the Jost
function (Eqs.(\ref{wron-jost1-csm}) and (\ref{wron-jost1-csm-dual})), we can also derive the relationship between
the Jost function in the original space and the Jost function in the dual space as follows.
\begin{enumerate}
\item When Eq.(\ref{KKt1})
  $\widetilde{\bvec{\mathcal{K}}}_\theta(E^*)=-\bvec{\mathcal{K}}_\theta^*(E)$
  is satisfied,
  \begin{eqnarray}
    \widetilde{\bvec{\Phi}}_{lj}^{(r)\theta *}(E^*)
    &=&
    (-)^l
    \bvec{\Phi}_{lj}^{(r)\theta}(E)
    \label{SolR1}
    \\
    \widetilde{\bvec{\Phi}}_{lj}^{(\pm)\theta *}(E^*)
    &=&
    (-)^l
    \bvec{\Phi}_{lj}^{(\pm)\theta}(E)
    \label{SolIR1}
  \end{eqnarray}
  The relation of the Jost function is obtained as
  \begin{eqnarray}
    \widetilde{\bvec{\mathcal{J}}}_{lj}^{(\pm)\theta *}(E^*)
    =
    \bvec{\mathcal{J}}_{lj}^{(\pm)\theta}(E)
    \label{JJt1}
  \end{eqnarray}
\item When Eq.(\ref{KKt2})
  $\widetilde{\bvec{\mathcal{K}}}_\theta(E^*)=
  \bvec{\sigma}_z\bvec{\mathcal{K}}_\theta^*(E)$
  is satisfied, 
  \begin{eqnarray}
    \widetilde{\bvec{\Phi}}_{lj}^{(r)\theta *}(E^*)
    &=&
    \bvec{\Phi}_{lj}^{(r)\theta}(E)
    \bvec{P}_+
    +
    (-)^l
    \bvec{\Phi}_{lj}^{(r)\theta}(E)
    \bvec{P}_-
    \label{SolR2}
    \\
    \widetilde{\bvec{\Phi}}_{lj}^{(\pm)\theta *}(E^*)
    &=&
    \bvec{\Phi}_{lj}^{(\mp)\theta}(E)
    \bvec{P}_+
    +
    (-)^l
    \bvec{\Phi}_{lj}^{(\pm)\theta}(E)
    \bvec{P}_-
    \label{SolIR2}
  \end{eqnarray}
  The relation of the Jost function is obtained as
  \begin{eqnarray}
    \widetilde{\bvec{\mathcal{J}}}_{lj}^{(\pm)\theta *}(E^*)
    &=&
    \bvec{P}_+\bvec{\mathcal{J}}_{lj}^{(\mp)\theta}(E)\bvec{P}_+
    \nonumber\\
    &&+
    \bvec{P}_-\bvec{\mathcal{J}}_{lj}^{(\pm)\theta}(E)\bvec{P}_-
    \nonumber\\
    &&+
    (-)^l\bvec{P}_+\bvec{\mathcal{J}}_{lj}^{(\pm)\theta}(E)\bvec{P}_-
    \nonumber\\
    &&+
    (-)^l\bvec{P}_-\bvec{\mathcal{J}}_{lj}^{(\mp)\theta}(E)\bvec{P}_+
    \nonumber
    \\
    &=&
    \begin{pmatrix}
      \bvec{\mathcal{J}}_{lj,11}^{(\mp)\theta}(E) &
      (-)^l\bvec{\mathcal{J}}_{lj,12}^{(\pm)\theta}(E) \\
      (-)^l\bvec{\mathcal{J}}_{lj,21}^{(\mp)\theta}(E) &
      \bvec{\mathcal{J}}_{lj,22}^{(\pm)\theta}(E)
    \end{pmatrix}
    \label{JJt2-2}
  \end{eqnarray}
\end{enumerate}
where $\bvec{P}_\pm$ are $2\times 2$ matrix defined by
\begin{eqnarray}
  \bvec{P}_+
  &\equiv&
  \begin{pmatrix}
    1 & 0 \\
    0 & 0
  \end{pmatrix}
  \mbox{ and }
  \bvec{P}_-
  \equiv
  \begin{pmatrix}
    0 & 0 \\
    0 & 1
  \end{pmatrix}.
\end{eqnarray}
Eq.(\ref{JJt1}) shows that if a pole $E_n$ of the S-matrix satisfying
$\det \bvec{\mathcal{J}}_{lj}^{(+)}(E_n)=0$ exists on the first Riemann sheet,
then there exists a pole on the first Riemann sheet in the dual space satisfying
$\det \widetilde{\bvec{\mathcal{J}}}_{lj}^{(+)}(E_n^*)=0$.

By using Eq.(\ref{JJt2-2}), we can obtain
\begin{eqnarray}
  &&
  \det\widetilde{\bvec{\mathcal{J}}}_{lj}^{(+)\theta *}(E^*)
  \nonumber\\
  &&=
  \bvec{\mathcal{J}}_{lj,11}^{(-)\theta}(E)
  \bvec{\mathcal{J}}_{lj,22}^{(+)\theta}(E)
  -
  \bvec{\mathcal{J}}_{lj,21}^{(-)\theta}(E)
  \bvec{\mathcal{J}}_{lj,12}^{(+)\theta}(E).
  \label{detJt}
\end{eqnarray}
When $E$ in $\det\bvec{\mathcal{J}}_{lj}^{(+)\theta}(E)$ lies on the first Riemann sheet of
the original space, the right-hand side of Eq.(\ref{detJt}) represents the expression
for the case where $E$ in $\det\bvec{\mathcal{J}}_{lj}^{(+)\theta}(E)$ lies on the second
Riemann sheet of the original space.
When a pole satisfying $\det\bvec{\mathcal{J}}_{lj}^{(+)\theta}(E_n)=0$ is found at
$E = E_n$ on the first Riemann sheet of the original space, there exists a pole at
$E=E_n^*$ on the second Riemann sheet; therefore, Eq.(\ref{detJt}) states that
when a pole satisfying $\det\bvec{\mathcal{J}}_{lj}^{(+)\theta}(E_n)=0$ is found at
$E=E_n$ on the first Riemann sheet of the original space, there also exists a pole
at $E=E_n$ on the second Riemann sheet of the dual space, satisfying
$\det\widetilde{\bvec{\mathcal{J}}}_{lj}^{(+)\theta *}(E_n)=0$. 
In other words, the first Riemann sheet of the original space and the second Riemann
sheet of the dual space share the same Riemann surface.


\subsection{Unitarity of the S-matrix}
To discuss the unitary nature of the S-matrix, we consider the Wronskian of the
flux-adjusted scattering wavefunction and its dual basis.
The Wronskian between the wave function of the original space and that of the dual
space is also a function independent of the coordinates, since substituting
$\bvec{\Phi}_{lj}^{(1)\theta\mathsf{T}}(E)$ with
$\widetilde{\bvec{\Phi}}_{lj}^{(1)\theta\dagger}(E^*)t$
in Eq.(\ref{identity}) yields the same result of zero on
the right-hand side.
The Wronskian between the scattering wave function in the original space and the
wave function in the dual space is zero due to the regularity near the origin
(i.e., $\bvec{\Psi}_{lj,F}^{(+)\theta}(r=0)=\widetilde{\bvec{\Psi}}_{lj,F}^{(+)\theta}(r=0)=0$),
that is
\begin{eqnarray}
  \bvec{\mathcal{W}}\left(\widetilde{\bvec{\Psi}}_{lj,F}^{(+)\theta *}(E^*),
  \bvec{\Psi}_{lj,F}^{(+)\theta}(E)\right)=0
  \label{wron-psipsit}
\end{eqnarray}
where
\begin{eqnarray}
  &&
  \bvec{\Psi}_{lj,F}^{(+)\theta\mathsf{T}}(E)
  \nonumber\\
  &&=
  \frac{1}{2}
  \left[
    \bvec{\Phi}_{lj,F}^{(-)\theta\mathsf{T}}(E)
    +
    \bvec{\mathcal{S}}_{lj,F}^{\theta}(E)
    \bvec{\Phi}_{lj,F}^{(+)\theta\mathsf{T}}(E)
    \right]
  \label{Scatwf-F}
  \\
  &&
  \widetilde{\bvec{\Psi}}_{lj,F}^{(+)\theta\mathsf{T}}(E)
  \nonumber\\
  &&=
  \frac{1}{2}
  \left[
    \widetilde{\bvec{\Phi}}_{lj,F}^{(-)\theta\mathsf{T}}(E)
    +
    \widetilde{\bvec{\mathcal{S}}}_{lj,F}^{\theta}(E)
    \widetilde{\bvec{\Phi}}_{lj,F}^{(+)\theta\mathsf{T}}(E)
    \right]
  \label{Scatwf-F-dual}
\end{eqnarray}
By inserting Eqs.(\ref{Scatwf-F}) and (\ref{Scatwf-F-dual}) into Eq.(\ref{wron-psipsit}),
we obtain
\begin{eqnarray}
  &&
  \widetilde{\bvec{\mathcal{S}}}_{lj,F}^{\theta\dagger}(E^*)
  \bvec{P}_+
  \bvec{\mathcal{S}}_{lj,F}^{\theta}(E)
  +
  (-)^l
  \widetilde{\bvec{\mathcal{S}}}_{lj,F}^{\theta\dagger}(E^*)
  \bvec{P}_-
  \nonumber\\
  &&
  -
  (-)^l
  \bvec{P}_-
  \bvec{\mathcal{S}}_{lj,F}^{\theta}(E)
  -
  \bvec{P}_+
  =0
\end{eqnarray}
when Eq.(\ref{KKt2})
$\widetilde{\bvec{\mathcal{K}}}_\theta(E^*)=\bvec{\sigma}_z\bvec{\mathcal{K}}_\theta^*(E)$
is satisfied.
Since $\bvec{P}_\pm$ has the properties like $\bvec{P}_+^2=\bvec{P}_\pm$,
$\bvec{P}_+\bvec{P}_-=0$ and $\bvec{P}_++\bvec{P}_-=\bvec{1}$, 
we can find
\begin{eqnarray}
  &&
  \bvec{P}_+
  \widetilde{\bvec{\mathcal{S}}}_{lj,F}^{\theta\dagger}(E^*)
  \bvec{P}_+
  \bvec{\mathcal{S}}_{lj,F}^{\theta}(E)
  \bvec{P}_+
  \nonumber\\
  &&=
  \bvec{P}_+
  \label{unitarity0}
  \\
  &&
  \bvec{P}_+
  \widetilde{\bvec{\mathcal{S}}}_{lj,F}^{\theta\dagger}(E^*)
  \bvec{P}_+
  \bvec{\mathcal{S}}_{lj,F}^{\theta}(E)
  \bvec{P}_-
  \nonumber\\
  &&=
  (-)^{l+1}
  \bvec{P}_+
  \widetilde{\bvec{\mathcal{S}}}_{lj,F}^{\theta\dagger}(E^*)
  \bvec{P}_-
\end{eqnarray}
and using these properties, we can derive
\begin{eqnarray}
  \bvec{P}_+
  \widetilde{\bvec{\mathcal{S}}}_{lj,F}^{\theta\dagger}(E^*)
  \bvec{P}_+
  \bvec{\Psi}_{lj,F}^{(+)\theta\mathsf{T}}(E)
  =
  \bvec{P}_+
  \widetilde{\bvec{\Psi}}_{lj,F}^{(+)\theta\dagger}(E^*)
  \label{unitarity0-wf}
\end{eqnarray}
If we express Eqs.(\ref{unitarity0}) and (\ref{unitarity0-wf}) in terms of matrix elements,
we get
\begin{eqnarray}
  \left(
  \widetilde{\bvec{\mathcal{S}}}_{lj,F}^{\theta\dagger}(E^*)
  \right)_{11}
  \left(
  \bvec{\mathcal{S}}_{lj,F}^{\theta}(E)
  \right)_{11}
  =1
  \label{unitarity-smat}
\end{eqnarray}
and
\begin{eqnarray}
  \left(
  \widetilde{\bvec{\mathcal{S}}}_{lj,F}^{\theta\dagger}(E^*)
  \right)_{11}
  \psi_{lj,F}^{(+1)\theta\mathsf{T}}(E)
  =
  \tilde{\psi}_{lj,F}^{(+1)\theta\dagger}(E^*)
  \label{unitarity-wf}
\end{eqnarray}
In scattering theory, it is known that a scattering wave function satisfying the
outgoing wave boundary condition and a scattering wave function satisfying the
incoming wave boundary condition are defined, and that these are related by a
unitary S-matrix.
Since $\psi_{lj,F}^{(+1)}$ behaves as a wave function satisfying the outgoing wave
scattering boundary condition in the positive energy region, and
$\tilde{\psi}_{lj,F}^{(+1)}$ satisfies the incoming wave scattering boundary condition,
Eq.(\ref{unitarity-wf}) indicates that
$\left(\widetilde{\bvec{\mathcal{S}}}_{lj,F}^{\theta\dagger}(E^*)\right)_{11}$ is an
S-matrix that satisfies the unitarity condition for the scattering problem
within the framework of the complex scaled HFB theory.

Applying the same discussion to the negative energy region (Re $E<0$),
the following relation is obtained.
\begin{eqnarray}
  \left(
  \widetilde{\bvec{\mathcal{S}}}_{lj,F}^{\theta\dagger}(E^*)
  \right)_{22}
  \left(
  \bvec{\mathcal{S}}_{lj,F}^{\theta}(E)
  \right)_{22}
  =1
  \label{unitarity-smat-neg}
\end{eqnarray}
and
\begin{eqnarray}
  \left(
  \widetilde{\bvec{\mathcal{S}}}_{lj,F}^{\theta\dagger}(E^*)
  \right)_{22}
  \psi_{lj,F}^{(+2)\theta\mathsf{T}}(E)
  =
  \tilde{\psi}_{lj,F}^{(+2)\theta\dagger}(E^*).
  \label{unitarity-wf-neg}
\end{eqnarray}

\begin{figure}[htbp]
  \includegraphics[width=\linewidth]{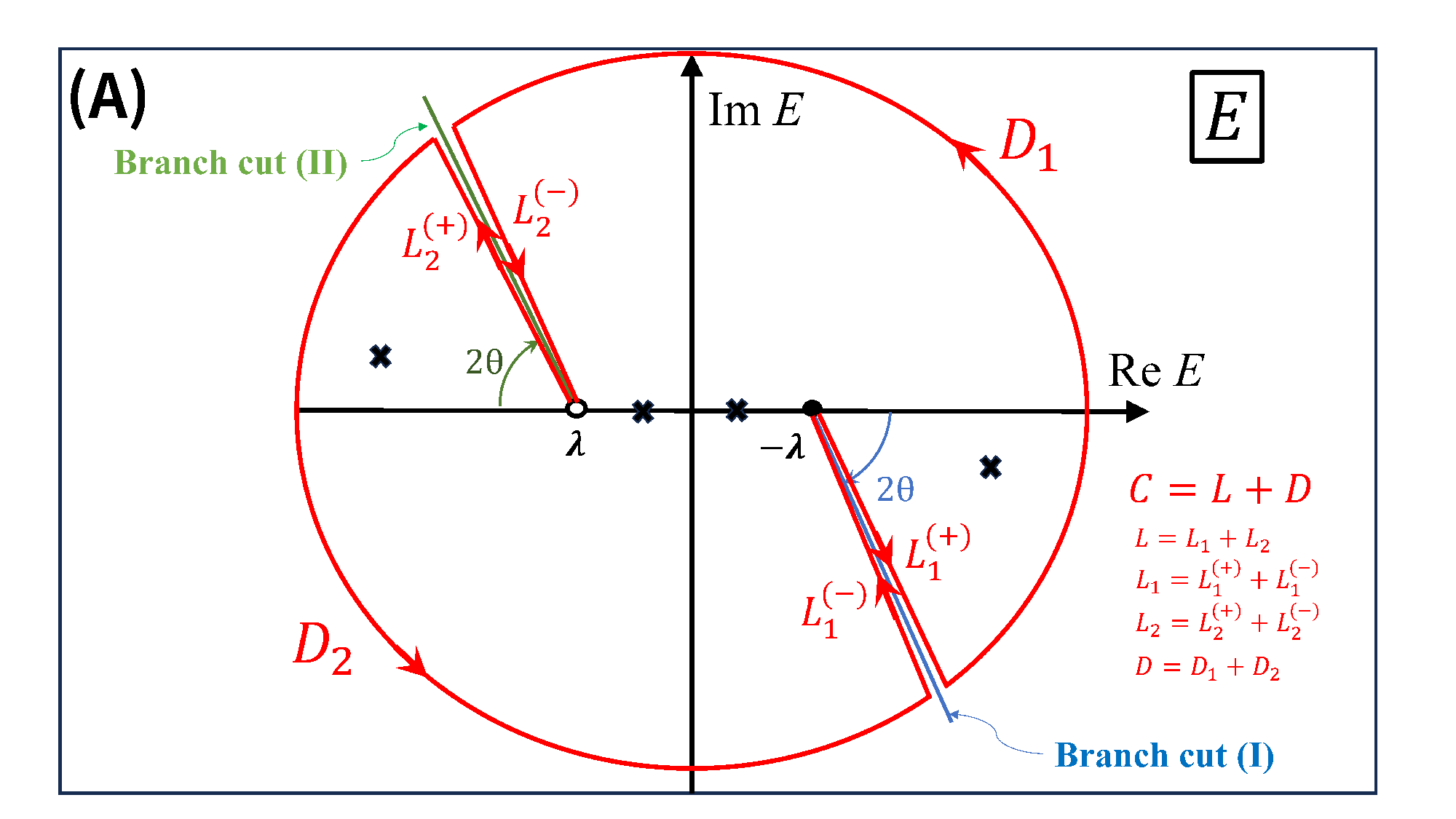}
  \includegraphics[width=\linewidth]{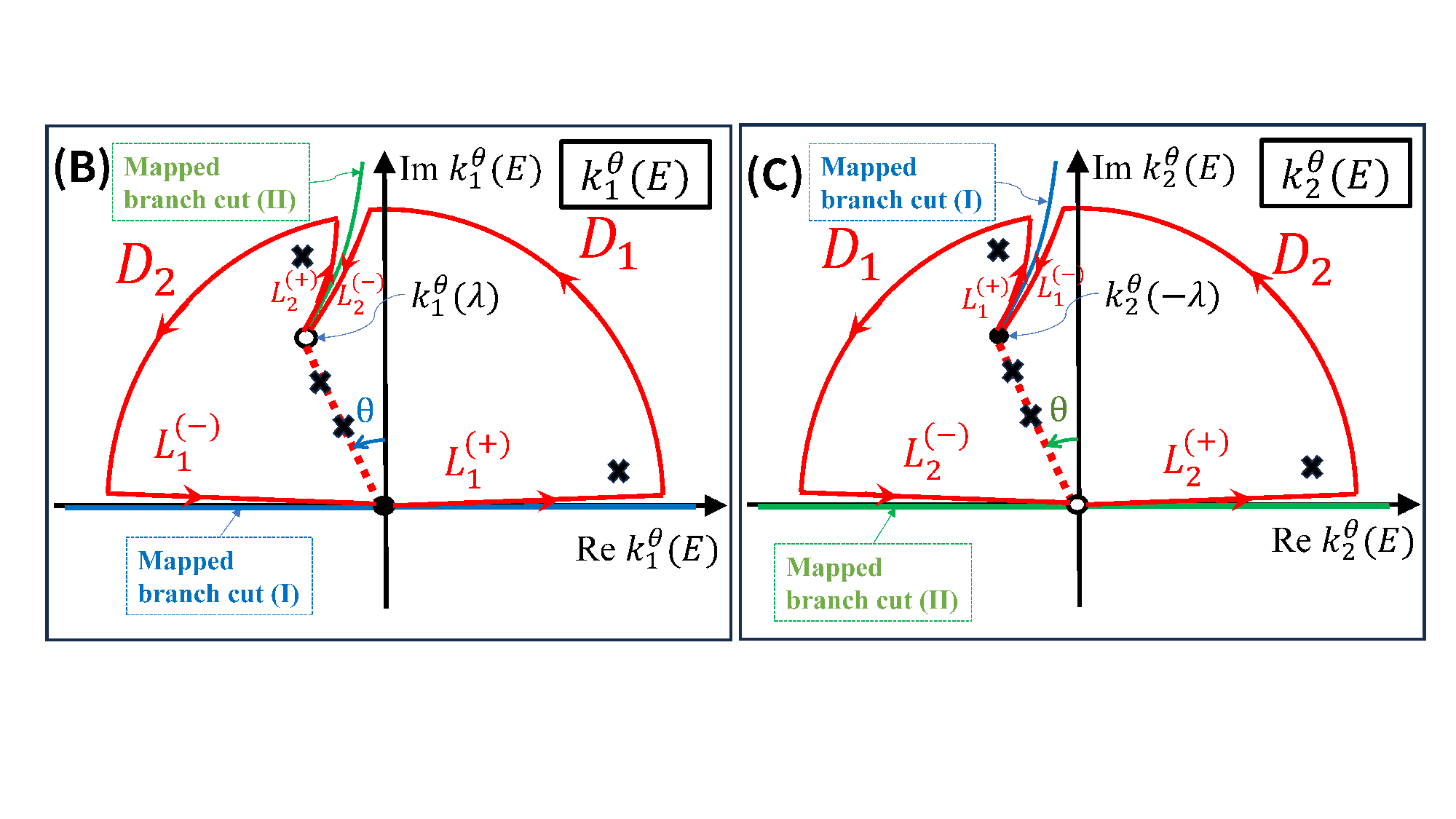}
  \caption{(Color online)
    Panel (A): Integration contour $C$ in the complex energy-$E$ plane. The contour $C$ consists of
    the paths $L = L_1 + L_2$ along the rotated branch cuts (I) and (II), and the outer
    arcs $D = D_1 + D_2$. The branch cuts are rotated by an angle $2\theta$ from the real axis.
    The crosses ($\times$) denote the poles of the $S$-matrix (or Green's function),
    where the poles in the fourth quadrant represent resonance states uncovered by the complex
    scaling. 
    Panels (B) and (C). Mapping of the integration contours and discrete poles onto the complex
    momentum planes $k_1^{\theta}(E)$ and $k_2^{\theta}(E)$. The figures illustrate the transformation
    of the branch cuts and the integration paths from the $E$-plane to the $k$-plane under complex
    scaling. The rotation angle $\theta$ in the momentum plane corresponds to the $2\theta$ scaling
    in the energy plane. Crosses ($\times$) denote the discrete eigenvalues, including bound and
    resonance states, captured within the deformed contours $L$ and $D$. The mapped branch cuts
    (I) and (II) are shown as blue and green lines, respectively. }
\label{Intpath}
\end{figure}

\subsection{Completeness and orthogonality}
The complex scaled HFB Green's function $\bvec{\mathcal{G}}^{\theta}_{lj}(r,r')$ is defined as
a function which satisfies 
\begin{eqnarray}
  \left[
    \frac{\hbar^2}{2m}
    \bvec{\mathcal{K}}_\theta^2(E)
    -
    \bvec{H}_\theta
    \right]
  \bvec{\mathcal{G}}^\theta_{lj}(r,r';E)
  =
  \delta(r-r')
  \bvec{\sigma}_z
  \label{HFBGreen},
\end{eqnarray}
and can be expressed as
\begin{eqnarray}
  &&
  \bvec{\mathcal{G}}_{lj}^{(+)\theta}(r,r')
  \nonumber\\
  &&=
  \frac{1}{i}
  \frac{2m}{\hbar^2}
  \left[
    \theta(r-r')
    \bvec{\Phi}_{lj}^{(+)\theta}(r)
    \bvec{\mathcal{K}}_\theta
    \bvec{\sigma}_z
    \bvec{\mathcal{J}}_{lj}^{(+)\theta-1}
    \bvec{\Phi}_{lj}^{(r)\theta\mathsf{T}}(r')
    \right.
    \nonumber\\
    &&
    \left.
    +
    \theta(r'-r)
    \bvec{\Phi}_{lj}^{(r)\theta}(r)
    \left(\bvec{\mathcal{J}}_{lj}^{(+)\theta\mathsf{T}}\right)^{-1}
    \bvec{\sigma}_z
    \bvec{\mathcal{K}}_\theta
    \bvec{\Phi}_{lj}^{(+)\theta\mathsf{T}}(r')
    \right]
  \nonumber\\
  \label{csmhfb_green}
\end{eqnarray}
Also, using this Green’s function, the scattering wave function
can be expressed as
\begin{eqnarray}
  &&
  \bvec{\Psi}_{lj}^{(+)\theta}(r;E)
  =
  \bvec{\chi}_{l}^{(r)\theta}(r;E)
  \nonumber\\
  &&
  +
  \int_0^\infty dr'
  \bvec{\mathcal{G}}_{lj}^{(+)\theta}(r,r')
  \bvec{\mathcal{V}}_\theta(r')
  \bvec{\chi}_{l}^{(r)\theta}(r';E)
  \label{LSeq}.
\end{eqnarray}

To obtain completeness, we consider the line integral along the closed path
$C (= L + D$) defined on the first Riemann sheet of the
complex energy $E$ plane, as shown in Fig.\ref{Intpath}, avoiding branch cuts.

By using the ``flux-adjusted'' solution matrix (Eq.(\ref{Phii-F})) and
S-matrix (Eq.(\ref{Smat-flux})), the Green function
Eq.(\ref{csmhfb_green}) can be rewritten as
\begin{eqnarray}
  \bvec{\mathcal{G}}_{lj}^{(+)\theta}(r,r')
  &=&
  \frac{1}{i}
  \frac{m}{\hbar^2}
  \left[
    \bvec{\Phi}_{lj,F}^{(+)\theta}(r)
    \bvec{\mathcal{S}}_{lj,F}^{\theta}
    \bvec{\Phi}_{lj,F}^{(+)\theta\mathsf{T}}(r')
    \right.
    \nonumber\\
    &&
    \left.
    +
    \theta(r-r')
    \bvec{\Phi}_{lj,F}^{(+)\theta}(r)
    \bvec{\Phi}_{lj,F}^{(-)\theta\mathsf{T}}(r')
    \right.
    \nonumber\\
    &&
    \left.
    +
    \theta(r'-r)
    \bvec{\Phi}_{lj,F}^{(-)\theta}(r)
    \bvec{\Phi}_{lj,F}^{(+)\theta\mathsf{T}}(r')
    \right]
  \label{csmhfb_green3}
\end{eqnarray}
Since $\bvec{\Phi}_{lj,F}^{(\pm)\theta}$ has no poles, the line integral of
a closed path $C$ on the first Riemann sheet of the Green’s function is
expressed as
\begin{eqnarray}
  &&
  \oint_C \frac{dE}{2\pi i}
  \bvec{\mathcal{G}}_{lj}^{(+)\theta}(r,r';E)
  \nonumber\\
  &&=
  \sum_n
  \bvec{\Phi}_{lj,F}^{(+)\theta}(r;E_n)
  \bvec{\mathcal{M}}_{nlj}^{\theta}
  \bvec{\Phi}_{lj,F}^{(+)\theta\mathsf{T}}(r';E_n)
  \label{gint1}
\end{eqnarray}
due to the residue theorem, where
\begin{eqnarray}
  &&
  \bvec{\mathcal{M}}_{nlj}^{\theta}
  \equiv
  \frac{1}{i}
  \frac{m}{\hbar^2}
  \sum_n
  \text{Res }[\bvec{\mathcal{S}}^{\theta}_{lj,F},E_n].
  \label{Mdef}
\end{eqnarray}
Since $\what{\bvec{\mathcal{S}}}^{\theta}_{lj}$ is a symmetric matrix
(Eq.(\ref{Smat_prop})), $\bvec{\mathcal{M}}_{nlj}^{\theta}$ is also a symmetric matrix.
It is known that complex symmetric matrices can be decomposed as
\begin{eqnarray}
  \bvec{\mathcal{M}}_{nlj}^{\theta}
  =
  \bvec{Q}^{\theta}\bvec{\Sigma}_n^\theta\bvec{Q}^{\theta\mathsf{T}}
  \label{takagidecom}
\end{eqnarray}
by the so-called Autonne-Takagi decomposition~\cite{Autonne1915,Takagi1925},
where $\bvec{Q}^\theta$ is a unitary matrix and $\bvec{\Sigma}_n^{\theta}$ is a diagonal matrix
with non-negative real singular values.
Autonne-Takagi decomposition is a special case of the Singular Value
Decomposition (SVD) for complex symmetric matrices.
Unlike the standard SVD, which uses two different unitary matrices, the
Takagi decomposition requires only one unitary matrix because of the
matrix's symmetry.

Normally, when performing a Takagi decomposition on a $2\times 2$ complex symmetric
matrix, the singular value matrix $\Sigma_n^\theta$ has two non-negative real singular
values as the diagonal components.
However, the defining Eq.(\ref{Mdef}) of $\bvec{\mathcal{M}}_{nlj}^{\theta}$
contains $\text{ adj }\bvec{\mathcal{J}}_{lj}^{(+)\theta}(E_n)$,
and since $\text{ adj }\bvec{\mathcal{J}}_{lj}^{(+)\theta}(E_n)$ is
a rank-1 matrix when $\det\bvec{\mathcal{J}}_{lj}^{(+)\theta}(E_n)=0$,
$\bvec{\mathcal{M}}_{nlj}^{\theta}$ is also a rank-1 matrix.
In this case, a rank-1 matrix always has only one non-zero singular value as
\begin{eqnarray}
  \bvec{\Sigma}_n^\theta
  =
  \begin{pmatrix}
    \sigma_n^\theta & 0 \\
    0 & 0
  \end{pmatrix}.
\end{eqnarray}
If we define ``the eigen wave function'' $\varphi_{nlj}$ for the pole $E_n$ as
\begin{eqnarray}
  &&
  \varphi_{nlj}^\theta(r)
  =
  \begin{pmatrix}
    \varphi_{nlj,1}^\theta(r) \\
    \varphi_{nlj,2}^\theta(r)
  \end{pmatrix}
  =
  \begin{pmatrix}
    \left(\bvec{\Phi}_{lj,F}^{(+)\theta}(r;E_n)
    \bvec{Q}^\theta
    \bvec{\Sigma}_n^{\theta\frac{1}{2}}
    \right)_{11}
    \\
    \left(\bvec{\Phi}_{lj,F}^{(+)\theta}(r;E_n)
    \bvec{Q}^\theta
    \bvec{\Sigma}_n^{\theta\frac{1}{2}}
    \right)_{21}
  \end{pmatrix}
  \nonumber\\
  &&=
  \sqrt{\sigma_n^\theta}
  \left[\phi_{lj,F}^{(+1)}(r;E_n)Q_{11}^\theta+\phi_{lj,F}^{(+2)}(r;E_n)Q_{21}^\theta\right],
  \label{eigenwf}
\end{eqnarray}
then Eq.(\ref{gint1}) can be rewritten as
\begin{eqnarray}
  \oint_C \frac{dE}{2\pi i}
  \bvec{\mathcal{G}}_{lj}^{(+)\theta}(r,r';E)
  =
  \sum_n
  \varphi_{nlj}^\theta(r)
  \varphi_{nlj}^{\theta\mathsf{T}}(r').
  \label{gint3}
\end{eqnarray}
The basis $\tilde{\varphi}_{nlj}$ representing poles existing on the
first Riemann sheet for the dual basis can similarly be defined.
Since it satisfies
$\bvec{H}_\theta^*\tilde{\varphi}_{nlj}
=\frac{\hbar^2}{2m}\widetilde{\bvec{\mathcal{K}}}_\theta^2(E_n^*)\tilde{\varphi}_{nlj}$
and has the relation $\tilde{\varphi}_{nlj}^*=\varphi_{nlj}$, Eq.(\ref{gint3})
can be expressed as
\begin{eqnarray}
  \oint_C \frac{dE}{2\pi i}
  \bvec{\mathcal{G}}_{lj}^{(+)\theta}(r,r';E)
  =
  \sum_n
  \varphi_{nlj}^\theta(r)
  \tilde{\varphi}_{nlj}^{\theta\dagger}(r')
  \label{gint-C}.
\end{eqnarray}
Next, we consider the integral of the path $L$ obtained by decomposing path $C$.
The $L$-path (consisting of $L_1$ and $L_2$) represents the integration contours
that run parallel to the branch cuts which are rotated by angle $2\theta$ due
to the complex scaling. As shown in Fig.\ref{Intpath},
let the path along branch cut (I) be denoted by $L_1$ and the path along
branch cut (II) by $L_2$.
When $E$ moves along the path $L_1^{(+)}$ in the complex energy plane,
$k_1^{(\theta)}$-plane, $k_1^{(\theta)}(E)$ moves on the real axis in the positive region;
when $E$ moves along the path $L_1^{(-)}$, $k_1^{(\theta)}$-plane,
$k_1^{(\theta)}(E)$ moves on the real axis in the negative region; that is,
$\arg k_1^{(\theta)}$ differs by $\pi$ between the paths $L_1^{(+)}$ and $L_1^{(-)}$.
On the other hand, the argument of $k_2^\theta$ does not change. 
Therefore, if we denote the energy on $L_1^{(+)}$ by $E_1^{(+)}$ and the energy on
$L_1^{(-)}$ by $E_1^{(-)}$, we have the relation for $\bvec{\mathcal{K}}_\theta(E)$ as
\begin{eqnarray}
  \bvec{\mathcal{K}}_\theta(E_1^{(-)})=-\bvec{\sigma}_z\bvec{\mathcal{K}}_\theta(E_1^{(+)})
\end{eqnarray}
By applying this relation to the flux-adjusted solutions, we obtain the following
symmetric properties as
\begin{eqnarray}
  \bvec{\Phi}_{lj,F}^{(r)\theta}(E_1^{(-)})
  &=&
  i(-)^l
  \bvec{\Phi}_{lj,F}^{(r)\theta}(E_1^{(+)})
  \bvec{P}_+
  +
  \bvec{\Phi}_{lj,F}^{(r)\theta}(E_1^{(+)})
  \bvec{P}_-
  \nonumber\\
  \\
  \bvec{\Phi}_{lj,F}^{(\pm)\theta}(E_1^{(-)})
  &=&
  i(-)^l
  \bvec{\Phi}_{lj,F}^{(\mp)\theta}(E_1^{(+)})
  \bvec{P}_+
  +
  \bvec{\Phi}_{lj,F}^{(\pm)\theta}(E_1^{(+)})
  \bvec{P}_-
  \nonumber\\
\end{eqnarray}
Applying these symmetrical properties and Eq.(\ref{unitarity-wf})
to the definition of the Green's function
Eq.(\ref{csmhfb_green3}), 
the integral along the path $L_1(=L_1^{(+)}+L_1^{(-)})$ of the Green's function is given by
\begin{eqnarray}
  &&
  \int_{L_1}\frac{dE}{2\pi i}
  \bvec{\mathcal{G}}_{lj}^{(+)\theta}(r,r';E)
  \nonumber\\
  &&
  =
  e^{-2i\theta}
  \int_{0}^{\infty}\frac{d|\lambda+E_1|}{2\pi i}
  \nonumber\\
  &&\times
  \left[
    \bvec{\mathcal{G}}_{lj}^{(+)\theta}(r,r';E_1^{(+)})
    -
    \bvec{\mathcal{G}}_{lj}^{(+)\theta}(r,r';E_1^{(-)})
    \right]
  \nonumber\\
  &&
  =
  -i
  \frac{4m}{\hbar^2}
  e^{-2i\theta}
  \int_{0}^{\infty}\frac{d|\lambda+E_1|}{2\pi i}
  \nonumber\\
  &&\hspace{10pt}
  \times
  \psi_{lj,F}^{(+1)\theta}(r;E_1^{(+)})
  \widetilde{\psi}_{lj,F}^{(+1)\theta\dagger}(r';E_1^{(+)*})
  \label{intL1-g0}
\end{eqnarray}
This integral can also be formally expressed as
\begin{eqnarray}
  &&
  \int_{L_1}\frac{dE}{2\pi i}
  \bvec{\mathcal{G}}_{lj}^{(+)\theta}(r,r';E)
  \nonumber\\
  &&
  =
  -
  \frac{2}{\pi}
  \int_{0}^{\infty}dk_1^{\theta}k_1^{\theta 2}
  \psi_{lj}^{(+1)\theta}(r;E_1^{(+)})
  \widetilde{\psi}_{lj}^{(+1)\theta\dagger}(r';E_1^{(+)*})
  \nonumber\\
  \label{intL1-g1}
\end{eqnarray}
using the scattering wave functions $\psi_{lj}^{(+1)\theta}$ and
$\widetilde{\psi}_{lj}^{(+1)\theta}$ defined by Eq.(\ref{defscat})
(not the flux-adjusted scattering wave functions
$\psi_{lj,F}^{(+1)\theta}$ or
$\widetilde{\psi}_{lj,F}^{(+1)\theta}$).
By performing a similar calculation for path $L_2$, we obtain the
following result.
\begin{eqnarray}
  &&
  \int_{L_2}\frac{dE}{2\pi i}
  \bvec{\mathcal{G}}_{lj}^{(+)\theta}(r,r';E)
  \nonumber\\
  &&
  =
  -
  \frac{2}{\pi}
  \int_{0}^{\infty}
  dk_2^{\theta}k_2^{\theta2}
  \psi_{lj}^{(+2)\theta}(r;E_2^{(+)})
  \widetilde{\psi}_{lj}^{(+2)\theta\dagger}(r';E_2^{(+)*})
  \nonumber\\
  \label{intL2-g1}
\end{eqnarray}
Combining Eqs.(\ref{intL1-g1}) and (\ref{intL2-g1}),
the integral with respect to the path $L(=L_1+L_2)$
is given by
\begin{eqnarray}
  &&
  \int_{L}\frac{dE}{2\pi i}
  \bvec{\mathcal{G}}_{lj}^{(+)\theta}(r,r';E)
  \nonumber\\
  &&
  =
  -
  \frac{2}{\pi}
  \sum_{i=1,2}
  \int_{0}^{\infty}dk_i^{\theta}k_i^{\theta 2}
  \psi_{lj}^{(+i)\theta}(r;E_i^{(+)})
  \widetilde{\psi}_{lj}^{(+i)\theta\dagger}(r';E_i^{(+)*}).
  \nonumber\\
  \label{gint-L}
\end{eqnarray}

The line integral of the circular path $D (=D_1+D_2)$ with infinite radius on the first
Riemann sheet of complex energy is zero unless $r=r'$, because the Green's function,
which is the integrand, is zero since Im $k_1>0$ and Im $k_2>0$.

When $r=r'$, the line integral of the path $D$ can be calculated as
\begin{eqnarray}
  &&
  \int_D \frac{dE}{2\pi i}
  \bvec{\mathcal{G}}_{lj}^{(+)\theta}(r,r;E)
  \nonumber\\
  &&
  =
  \int_D \frac{dE}{2\pi i}
  \frac{1}{i}
  \frac{m}{\hbar^2}
  \left[
    \bvec{\Phi}_{lj,F}^{(+)\theta}(r)
    \bvec{\Phi}_{lj,F}^{(-)\theta\mathsf{T}}(r)
    \right]
  \nonumber\\
  &&
  =
  \lim_{|k_1^\theta|\to\infty}
  \frac{|k_1^\theta|}{2\pi i}
  \int_0^{\pi} e^{i\theta}d\theta
  \begin{pmatrix}
    1 & 0 \\
    0 & 0
  \end{pmatrix}
  \nonumber\\
  &&
  -
  \lim_{|k_2^\theta|\to\infty}
  \frac{|k_2^\theta|}{2\pi i}
  \left(\int_{\pi/2}^\pi+\int_0^{\pi/2}\right)
  e^{i\theta}d\theta
  \begin{pmatrix}
    0 & 0 \\
    0 & -1
  \end{pmatrix}
  \nonumber\\
  &&
  =
  \begin{pmatrix}
    1 & 0 \\
    0 & 1
  \end{pmatrix}
  \lim_{|k|\to\infty}
  \frac{|k|}{\pi}
  \to
  \bvec{1}\infty
\end{eqnarray}
However, since $\infty$ in complex analysis does not denote
infinity in the positive real direction, the result can be
expressed as follows using an arbitrary complex number $C$.
\begin{eqnarray}
  \int_D \frac{dE}{2\pi i}
  \bvec{\mathcal{G}}_{lj}^{(+)\theta}(r,r';E)
  =
  \bvec{1}C\delta(r-r').
  \label{gint-D}
\end{eqnarray}
By inserting Eqs.(\ref{gint-C}), (\ref{gint-L}) and (\ref{gint-D})
into the complex contour
integral identity $\oint_C-\int_L=\int_D$, the following relationship is obtained;
\begin{eqnarray}
  &&
  \sum_n
  \varphi_{nlj}^\theta(r)
  \tilde{\varphi}_{nlj}^{\theta\dagger}(r')
  \nonumber\\
  &&
  +
  \frac{2}{\pi}
  \sum_{i=1,2}
  \int_{0}^{\infty}dk_i^{\theta}k_i^{\theta 2}
  \psi_{lj}^{(+i)\theta}(r;E_i^{(+)})
  \widetilde{\psi}_{lj}^{(+i)\theta\dagger}(r';E_i^{(+)*})
  \nonumber\\
  &&
  =
  C\bvec{1}\delta(r-r').
  \label{completeness}
\end{eqnarray}
This is a so-called completeness relation, but $C$ is still
undetermined here. 

The orthogonality of shown below can be proven using Green's theorem based on
the regularity of both wave functions near the origin $r=0$ and the asymptotic
behavior of $\varphi_{nlj}$ or $\tilde{\varphi}_{nlj}$ converging to
zero at infinity ($r=\infty$).
\begin{eqnarray}
  \int_0^\infty dr
  \widetilde{\varphi}_{nlj}^{\dagger}(r)
  \psi_{lj}^{(+i)\theta}(r;E)
  &=&
  \int_0^\infty dr
  \widetilde{\psi}_{lj}^{(+i)\theta\dagger}(r;E)
  \varphi_{nlj}(r)
  \nonumber\\
  &=&0 \text{ (for $i\in 1,2$)}
  \label{orthphipsi}
\end{eqnarray}
However, it is important to note that this orthogonality holds
only if the condition Im $k_1^\theta(E) <$ Im $k_1^\theta(E_n)$ is
satisfied. 

Applying the Green's theorem to the l.h.s of the following identity
that $\varphi_{nlj}^\theta$ and $\tilde{\varphi}^\theta_{nlj}$ satisfy the
following identity;
\begin{eqnarray}
  &&
  \int_0^{\infty} dr
  \tilde{\varphi}^{\theta\dagger}_{nlj}
  \left[
    \bvec{H}^{\dagger}
    \bvec{\sigma}_z
    -
    \bvec{\sigma}_z
    \bvec{H}_{\theta}
    \right]
  \varphi_{n'lj}^\theta
  \nonumber\\
  &&
  =
  (E_{n}-E_{n'})e^{2i\theta}
  \int_0^{\infty} dr
  \tilde{\varphi}^{\theta\dagger}_{nlj}
  \varphi_{n'lj}^\theta,
  \label{identity-varphi}
\end{eqnarray}
we can obtain
\begin{eqnarray}
  (E_{n}-E_{n'})
  e^{2i\theta}
  \int_0^{\infty} dr
  \tilde{\varphi}^{\theta\dagger}_{nlj}
  \varphi_{n'lj}^\theta
  =0
  \label{identity-varphi2}
\end{eqnarray}
because of the regularity of both wave functions near the origin
$r=0$ and the asymptotic behavior of both wave functions converging
to zero at infinity ($r=\infty$). 
This result implies that the orthogonality of
$\varphi_{nlj}^\theta$ and $\tilde{\varphi}^\theta_{nlj}$ is given by
\begin{eqnarray}
  e^{2i\theta}
  \int_0^{\infty} dr
  \tilde{\varphi}^{\theta\dagger}_{nlj}
  \varphi_{n'lj}^\theta
  =\delta_{nn'}.
  \label{normal}
\end{eqnarray}
However, this orthogonality holds only when the pole is found on
the first Riemann sheet and the wave function converges to zero
at infinity. 

Applying Eqs.(\ref{orthphipsi}) and (\ref{normal}) to
Eq.(\ref{completeness}), we can obtain $C=e^{-2i\theta}$,
finally we can obtain the completeness as
\begin{eqnarray}
  &&
  \sum_n
  \varphi_{nlj}^\theta(r)
  \tilde{\varphi}_{nlj}^{\theta\dagger}(r')
  \nonumber\\
  &&
  +
  \frac{2}{\pi}
  \sum_{i=1,2}
  \int_{0}^{\infty}dk_i^{\theta}k_i^{\theta 2}
  \psi_{lj}^{(+i)\theta}(r;E_i^{(+)})
  \widetilde{\psi}_{lj}^{(+i)\theta\dagger}(r';E_i^{(+)*})
  \nonumber\\
  &&
  =
  e^{-2i\theta}\bvec{1}\delta(r-r').
  \label{completeness2}
\end{eqnarray}
By using Eqs.(\ref{orthphipsi}) and (\ref{completeness2}),
we can derive the orthogonality for $\psi_{lj}^{(+i)\theta}$ as
\begin{eqnarray}
  &&
  \frac{2}{\pi}
  e^{2i\theta}
  \int_0^\infty dr
  \widetilde{\psi}_{lj}^{(+i)\theta\dagger}(r;E^*_i)
  \psi_{lj}^{(+i')\theta}(r;E)
  \nonumber\\
  &&\hspace{30pt}
  =
  \delta_{ii'}
  \frac{\delta(k_i^{\theta}(E_i)-k_i^{\theta}(E))}
       {k_i^{\theta}(E_i)k_i^{\theta}(E)}.
\end{eqnarray}
If we assume that the Green’s function can be expanded in terms
of the basis appearing in Eq.(\ref{completeness2}),
we can obtain the so-called ``spectral representation'' of the
Green’s function as
\begin{eqnarray}
  &&
  \bvec{\mathcal{G}}_{lj}^{(+)\theta}(r,r';E)
  \nonumber\\
  &&=
  \sum_{n>0}
  \left[
    \frac{
      \varphi_{nlj}^\theta(r)
      \tilde{\varphi}_{nlj}^{\theta\dagger}(r')
      }{E-E_{nlj}}
      +
      \frac{
        \varphi_{\overline{nlj}}^\theta(r)
        \tilde{\varphi}_{\overline{nlj}}^{\theta\dagger}(r')
      }{E+E_{nlj}}
      \right]
  \nonumber\\
  &&
  +
  \frac{2}{\pi}
  \int_{0}^{\infty}dk_1^{\theta}(E_1)(k_1^{\theta}(E_1))^2
  \nonumber\\
  &&\hspace{10pt}
  \times
  \left[
    \frac{
      \psi_{lj}^{(+1)\theta}(r;E_1^{(+)})
      \widetilde{\psi}_{lj}^{(+1)\theta\dagger}(r';E_1^{(+)*})
    }{E-E_1}
    \right.
    \nonumber\\
    &&\hspace{15pt}
    \left.
    +
    \frac{
      \psi_{lj}^{(+2)\theta}(r;-E_1^{(+)})
      \widetilde{\psi}_{lj}^{(+2)\theta\dagger}(r';-E_1^{(+)*})
      }{E+E_1}
    \right]
  \label{GreenSP}
\end{eqnarray}
where $\varphi_{\overline{nlj}}^\theta$ is a negative-energy solution
that satisfies the symmetry given by Eq.(\ref{detjost-neg}).
If we express the first term on the left-hand side of the
completeness relation in Eq.(\ref{completeness2})
using $\varphi_{\overline{nlj}}^\theta$, it can be expressed as
\begin{eqnarray}
  &&
  \sum_n
  \varphi_{nlj}^\theta(r)
  \tilde{\varphi}_{nlj}^{\theta\dagger}(r')
  \nonumber\\
  &&
  =
  \sum_{n>0}
  \varphi_{nlj}^\theta(r)
  \tilde{\varphi}_{nlj}^{\theta\dagger}(r')
  +
  \sum_{n<0}
  \varphi_{\overline{nlj}}^\theta(r)
  \tilde{\varphi}_{\overline{nlj}}^{\theta\dagger}(r').
\end{eqnarray}
By multiplying by
$
\left[
  \frac{\hbar^2}{2m}
  \bvec{\mathcal{K}}_\theta^2(E)
  -
  \bvec{H}_\theta
  \right]
$ on the left-hand side of Eq.(\ref{GreenSP}),
it is easy to confirm that the Green’s function expressed by
Eq.(\ref{GreenSP}) satisfies the definition of the Green’s
function Eq.(\ref{HFBGreen}), due to the existence of
completeness Eq.(\ref{completeness2}).

\subsection{T-matrix and its residue}
Since $\left(\bvec{\mathcal{S}}_{lj,F}^{\theta}(E)\right)_{11}$ is the S-matrix for a scattering
problem that satisfies the unitarity condition as shown in Eq.(\ref{unitarity-smat}) within the
theoretical framework of this paper, the T-matrix is defined as
\begin{eqnarray}
  T_{lj}^{\theta}(E)
  \equiv
  \frac{i}{2}
  \left[
    \left(\bvec{\mathcal{S}}_{lj,F}^{\theta}(E)\right)_{11}-1
    \right]
  \label{Tmatdef}
\end{eqnarray}
where the S-matrix can be expressed by using the Jost function as
\begin{eqnarray}
  &&
  \left(\bvec{\mathcal{S}}_{lj,F}^{\theta}(E)\right)_{11}
  \nonumber\\
  &&=
  \frac{
  \bvec{\mathcal{J}}_{lj,11}^{(-)\theta}(E)
  \bvec{\mathcal{J}}_{lj,22}^{(+)\theta}(E)
  -
  \bvec{\mathcal{J}}_{lj,21}^{(-)\theta}(E)
  \bvec{\mathcal{J}}_{lj,12}^{(+)\theta}(E).
  }{
  \bvec{\mathcal{J}}_{lj,11}^{(+)\theta}(E)
  \bvec{\mathcal{J}}_{lj,22}^{(+)\theta}(E)
  -
  \bvec{\mathcal{J}}_{lj,21}^{(+)\theta}(E)
  \bvec{\mathcal{J}}_{lj,12}^{(+)\theta}(E).
  }.
  \nonumber\\
  \label{S11det}
\end{eqnarray}
Due to the Mittag-Leffler theorem~\cite{Rakityansky,Mizuyama2025},
the T-matrix can be decomposed into a pole contribution and other terms as
\begin{eqnarray}
  T_{lj}^{\theta}(E)
  &=&
  T_{lj,reg}^{\theta}(E)
  +
  \sum_n
  \frac{\operatorname{Res}[T_{lj}^{\theta},E_{nlj}]}{E-E_{nlj}}.
  \label{MLtheorem}
\end{eqnarray}
where $\operatorname{Res}[T_{lj}^{\theta},E_{nlj}]$ is the residue for a pole $E_n$
and is calculated as
\begin{eqnarray}
  \operatorname{Res}[T_{lj}^{\theta},E_{nlj}]
  =
  \oint_{C_n}\frac{dE}{2\pi i}
  T_{lj}^{\theta}(E).
  \label{resT}
\end{eqnarray}
By using the integral representation of the Jost function Eq.(\ref{jost-int}) and
the definition of the scattering wave function Eq.(\ref{defscat}), Eq.(\ref{Tmatdef}) can be
rewritten as
\begin{eqnarray}
  T_{lj}^{\theta}(E)
  &=&
  \frac{2mk_1^\theta(E)}{\hbar^2}
  \int_0^\infty dr
  \chi_l^{(r1)\theta\mathsf{T}}(E)
  \bvec{\mathcal{V}}_\theta
  \psi_{lj}^{(+1)\theta}(E).
  \nonumber\\
  \label{Tmat-int}
\end{eqnarray}
Since $\psi_{lj}^{(+1)\theta}(E)$ satisfies Eq.(\ref{LSeq}), we can obtain
\begin{eqnarray}
  &&
  T_{lj}^{\theta}(E)
  \nonumber\\
  &&=
  \frac{2mk_1^\theta(E)}{\hbar^2}
  \int_0^\infty dr
  \chi_l^{(r1)\theta\mathsf{T}}(E)
  \bvec{\mathcal{V}}_\theta
  \chi_l^{(r1)\theta\mathsf{T}}(E)
  \nonumber\\
  &&
  +
  \frac{2mk_1^\theta(E)}{\hbar^2}
  \int\int_0^\infty drdr'
  \nonumber\\
  &&\times
  \chi_l^{(r1)\theta\mathsf{T}}(r;E)
  \bvec{\mathcal{V}}_\theta
  \bvec{\mathcal{G}}_{lj}^{(+)\theta}(r,r')
  \bvec{\mathcal{V}}_\theta
  \chi_l^{(r1)\theta}(r';E).
  \nonumber\\
  \label{Tmat-int2}
\end{eqnarray}
Applying the spectral representation of the Green’s function Eq.(\ref{GreenSP})
to the T-matrix, since the first term on the right-hand side of Eq.(\ref{Tmat-int2})
has no poles, the residues of the T-matrix can be expressed as
\begin{eqnarray}
  \operatorname{Res}[T_{lj}^{\theta},E_{nlj}]
  =
  \gamma_n^2
  \label{resT2}
\end{eqnarray}
where
\begin{eqnarray}
  &&
  \gamma_n
  \equiv
  \sqrt{\frac{2mk_1^\theta(E_{nlj})}{\hbar^2}}
  \nonumber\\
  &&\hspace{10pt}
  \times
  \int_0^\infty dr
  \chi_l^{(r1)\theta\mathsf{T}}(r;E_{nlj})
  \bvec{\mathcal{V}}_\theta(r)
  \varphi_{nlj}^{\theta}(r).
  \label{gamma}
\end{eqnarray}
$\gamma_n$ can be decomposed into two terms;
a scattering term by  $\gamma_n^{(1)}$ and
a pairing potential scattering term $\gamma_n^{(2)}$ as
\begin{eqnarray}
  &&
  \gamma_n
  =\gamma_n^{(1)}+\gamma_n^{(2)}
  \label{gamma-sep}
  \\
  &&
  \gamma_n^{(1)}
  \equiv
  \sqrt{\frac{2mk_1^\theta(E_{nlj})}{\hbar^2}}
  \nonumber\\
  &&\hspace{10pt}
  \times
  \int_0^\infty dr
  F_l(k_1^\theta r)
  e^{2i\theta}U_{lj}(r e^{i\theta})
  \varphi_{nlj,1}^{\theta}(r).
  \label{gamma1}
  \\
  &&
  \gamma_n^{(2)}
  \equiv
  \sqrt{\frac{2mk_1^\theta(E_{nlj})}{\hbar^2}}
  \nonumber\\
  &&\hspace{10pt}
  \times
  \int_0^\infty dr
  F_l(k_1^\theta r)
  e^{2i\theta}\Delta_{lj}(r e^{i\theta})
  \varphi_{nlj,2}^{\theta}(r).
  \label{gamma2}
\end{eqnarray}
Therefore, the second term on the right-hand side of Eq.(\ref{MLtheorem}) can be decomposed as
\begin{eqnarray}
  &&
  \sum_n
  \frac{\operatorname{Res}[T_{lj}^{\theta},E_{nlj}]}{E-E_{nlj}}.
  =
  \sum_n
  \frac{\gamma_n^2}{E-E_{nlj}}.
  \nonumber\\
  &&=
  \sum_n
  \frac{\gamma_n^{(1)2}}{E-E_{nlj}}.
  +
  2
  \sum_n
  \frac{\gamma_n^{(2)}\gamma_n^{(2)}}{E-E_{nlj}}.
  \nonumber\\
  &&
  +
  \sum_n
  \frac{\gamma_n^{(2)2}}{E-E_{nlj}}.
  \nonumber\\
  &&=
  t^{(n)}_1(E)+t^{(n)}_2(E)+t^{(n)}_3(E).
  \label{resTmat}
\end{eqnarray}
The first and third terms represent the contributions from mean-field potential scattering
and pairing potential scattering at the poles of the T-matrix, respectively.
The second term represents the contribution from their interference. 

\section{Numerical results and discussion}

\subsection{Numerical setup}
We adopt the Woods-Saxon potential given by
\begin{eqnarray}
  U_{lj}(r)
  =
  V_0 f_{WS}(r)+\frac{\hbar^2l(l+1)}{2mr^2}
  +
  V_{ls}\bvec{l}\idot\bvec{s}\frac{1}{r}\frac{df_{WS}(r)}{dr}
  \label{WSpot}
\end{eqnarray}
for the mean field $U_{lj}$(r), where
\begin{eqnarray}
  f_{WS}(r)
  =
  \frac{1}{1+\exp\left(\frac{r-R}{a}\right)}
\end{eqnarray}
with the standard parameters;
$V_0=-51$ MeV, $a=0.67$ fm, $V_{ls}=-18$ MeV fm$^2$ and $R=r_0A^{\frac{1}{3}}$ with
$r_0=1.27$ fm and the value of $A$ is set as $A=24$ assuming 
neutron scattering targeting nuclei near the dripline in the oxygen
neighborhood of medium-mass nuclei.
We also set the chemical potential $\lambda$ to $\lambda=-1.0$ with
the same reasoning.

We adopt the volume-type pairing potential which is given by
\begin{eqnarray}
  \Delta(r)=V_{pair}f_{WS}(r)
\end{eqnarray}
where the pairing potential strength $V_{pair}$ is set to reproduce
the given average pairing gap $\bra\Delta\ket$ using
\begin{eqnarray}
  \bra\Delta\ket=\frac{\int_0^\infty dr r^2\Delta(r) f_{WS}(r)}{\int_0^\infty dr r^2 f_{WS}(r)}.
\end{eqnarray}
In this paper, we used the Numerov method to solve the HFB equation, which is given by a
system of second-order differential equations, in the region $r < 20$ fm with a mesh size
of $\Delta r = 0.1$ fm. Furthermore, we employed the Newton-Raphson method to find the
pole defined by $\det\bvec{\mathcal{J}}_{lj}^{(+)\theta}(E_n)=0$. 

\subsection{Systematic verification of the $\theta$-invariance and normalization}
To systematically investigate the properties of the quasiparticle resonance states and to verify the consistency of the proposed complex-scaled Jost-HFB framework, the
numerical results are presented in three stages, summarized in Tables \ref{table1}, \ref{table2}, and \ref{table3}.

\begin{table*}
  \caption{
    Calculated single-particle energies $\epsilon_{nlj}$ and quasiparticle resonance energies $E_{nlj}$ without pairing correlation ($\langle \Delta \rangle = 0.0$ MeV)
    and complex scaling ($\theta =0.0$). The chemical potential is fixed at $\lambda = -1.0$ MeV. The states are classified as particle-type, hole-type, and shape resonances.
    The last column represents the normalization of the quasiparticle wave functions according to Eq.(\ref{normal}). At $\theta = 0.0$, only the bound-state-like particle-type
    quasiparticles are integrable and normalized to 1.00, while others are divergent or non-integrable (indicated by "--").
  }
  \label{table1}
  \begin{ruledtabular}
    \begin{tabular}{cccccccccc}
      & & & \multicolumn{7}{c}{$\lambda=-1.0$ MeV, $\bra\Delta\ket=0.0$ MeV, $\theta=0.0$}\\
      \cline{4-10}
      No. & $nlj$ & $\epsilon_{nlj}$
      & $E_{nlj}$ & $k_1^{\theta}(E_{nlj})$ & $k_2^{\theta}(E_{nlj})$
      & \multicolumn{1}{c}{$Q^{\theta}_{11}$} & \multicolumn{1}{c}{$Q^{\theta}_{21}$} & $\sigma^{\theta}_{nlj}$ 
      & \makecell[l]{$e^{+2i\theta}$\\$\times\bra\widetilde{\varphi}_{nlj}^{\theta}|\varphi_{nlj}^{\theta}\ket$}
      \\
      & & [MeV] & [MeV] & [fm$^{-1}$] & [fm$^{-1}$] & \multicolumn{1}{c}{[1]} & \multicolumn{1}{c}{[1]} & [fm$^{-2}$] & [1] \\
      \colrule
      \multicolumn{10}{c}{--particle-type quasiparticle resonance--}\\
      (1) & $2p_{3/2}$ & $ -0.21$  & $ 0.79$ & $i 0.10$ & $i 0.29$ & $-0.71-i 0.71$ & $ 0.00$        & $7.56\times 10^{-3}$ & $1.00$  \\
      (2) & $1f_{7/2}$ & $ -0.93$  & $ 0.07$ & $i 0.21$ & $i 0.23$ & $-0.71-i 0.71$ & $ 0.00$        & $3.20\times 10^{-4}$ & $1.00$  \\
      \multicolumn{10}{c}{--hole-type quasiparticle resonance--}\\
      (3) & $1d_{3/2}$ & $ -5.09$  & $ 4.09$ & $0.39$   & $i 0.50$ & $0.00$          & $ 0.69+i 0.73$ & $9.81\times 10^{-1}$ & $--$   \\
      (4) & $2s_{1/2}$ & $ -8.62$  & $ 7.62$ & $0.57$   & $i 0.64$ & $0.00$          & $-0.98-i 0.21$ & $2.78\times 10^{1}$  & $--$   \\
      (5) & $1d_{5/2}$ & $-12.17$  & $11.17$ & $0.70$   & $i 0.77$ & $0.00$          & $-0.82-i 0.58$ & $3.12\times 10^{1}$  & $--$   \\
      (6) & $1p_{1/2}$ & $-19.65$  & $18.65$ & $0.92$   & $i 0.97$ & $0.00$          & $ 0.70-i 0.71$ & $2.76\times 10^{2}$  & $--$   \\
      (7) & $1p_{3/2}$ & $-23.61$  & $22.62$ & $1.02$   & $i 1.07$ & $0.00$          & $ 0.67-i 0.75$ & $7.91\times 10^{2}$  & $--$   \\
      (8) & $1s_{1/2}$ & $-34.78$  & $33.78$ & $1.26$   & $i 1.30$ & $0.00$          & $ 0.60+i 0.80$ & $2.57\times 10^{3}$  & $--$   \\
      \multicolumn{10}{c}{--Shape resonance--}\\
      (9) & $f_{5/2}$ & $--$  & $7.75-i1.63$ & $0.57-i0.07$   & $0.06+i 0.65$ & $0.51+i0.86$ & $ 0.00$ & $6.86\times 10^{-2}$  & $--$   \\
      (10)& $g_{9/2}$ & $--$  & $10.06-i 1.14$ & $0.66-i0.04$   & $0.04+i 0.73$ & $0.34+i0.94$ & $ 0.00$ & $5.17\times 10^{-2}$  & $--$ 
    \end{tabular}
  \end{ruledtabular}
\end{table*}

In these tables, the first three columns list the index (No.), the angular momentum quantum numbers ($nlj$), and the single-particle energies ($\epsilon_{nlj}$) of the
Woods-Saxon mean-field potential $U_{lj}(r)$ [Eq. (\ref{WSpot})]. The complex quasiparticle energy $E_{nlj}$ represents the position of the resonance pole in the complex energy
plane, where the imaginary part corresponds to half of the decay width. The momenta $k_1^\theta(E_{nlj})$ and $k_2^\theta(E_{nlj})$ are the complex-scaled momenta for
the upper and lower components, respectively, as defined by Eqs. (\ref{defk1-csm}) and (\ref{defk2-csm}). The coefficients $Q_{11}^\theta$ and $Q_{21}^\theta$ correspond to the elements of the
unitary matrix $Q^\theta$ obtained through the ``Autonne-Takagi factorization'' of the residue matrix $\mathcal{M}_{nlj}^\theta$ [Eq. (\ref{takagidecom})], representing the mixing of
the upper and lower components of the quasiparticle wave function at the resonance pole. The singular value $\sigma_{nlj}^\theta$ denotes the magnitude of the residue
as defined in Eq. (101). The final column shows the result of the normalization integral based on the dual basis defined by Eq.~(\ref{normal}).

\begin{table*}
  \caption{
    Numerical results with the pairing correlation $\langle \Delta \rangle = 3.0$ MeV at $\theta = 0.0$.
    Due to the pairing interaction, the single-particle levels in Table \ref{table1} are transformed into quasiparticle resonances in the complex energy plane.
    Consequently, all wave functions become divergent at $r \to \infty$ for $\theta = 0.0$, leading to the undefined normalization ("--") in the last column for all states.
  }
  \label{table2}
  \begin{ruledtabular}
    \begin{tabular}{cccccccccc}
      & & & \multicolumn{7}{c}{$\lambda=-1.0$ MeV, $\bra\Delta\ket=3.0$ MeV, $\theta=0.0$}\\
      \cline{4-10}
      No. & $nlj$ & $\epsilon_{nlj}$
      & $E_{nlj}$ & $k_1^{\theta}(E_{nlj})$ & $k_2^{\theta}(E_{nlj})$
      & \multicolumn{1}{c}{$Q^{\theta}_{11}$} & \multicolumn{1}{c}{$Q^{\theta}_{21}$} & $\sigma^{\theta}_{nlj}$ 
      & \makecell[l]{$e^{+2i\theta}$\\$\times\bra\widetilde{\varphi}_{nlj}^{\theta}|\varphi_{nlj}^{\theta}\ket$}
      \\
      & & [MeV] & [MeV] & [fm$^{-1}$] & [fm$^{-1}$] & \multicolumn{1}{c}{[1]} & \multicolumn{1}{c}{[1]} & [fm$^{-2}$] & [1] \\
      \colrule
      \multicolumn{10}{c}{--particle-type quasiparticle resonance--}\\
      (1) & $2p_{3/2}$ & $ -0.21$  & $ 1.35-i 0.17 $ & $0.13-i 0.03$ & $0.01 +i0.34$ & $0.11+i0.13$ & $-0.39+i0.91$ & $2.50\times 10^{-1}$  & $--$ \\
      (2) & $1f_{7/2}$ & $ -0.93$  & $ 2.45-i 0.01 $ & $0.26-i 0.00$ & $0.00+i0.41$  & $ 0.00+i0.16$ & $ 0.69-i0.71$ & $2.17\times 10^{-2}$ & $--$ \\
      \multicolumn{10}{c}{--hole-type quasiparticle resonance--}\\
      (3) & $1d_{3/2}$ & $ -5.09$  & $ 4.94-i 0.25 $ & $0.44-i 0.01$ & $0.01 +i0.54$ & $ 0.01+i0.08$ & $-0.76-i0.64$ & $1.64$             &   $--$\\
      (4) & $2s_{1/2}$ & $ -8.62$  & $ 8.22-i 0.48 $ & $0.59-i 0.02$ & $0.02 +i0.67$ & $-0.01+i0.01$ & $-0.73-i0.68$ & $8.70\times 10^{1}$ &  $--$\\
      (5) & $1d_{5/2}$ & $-12.17$  & $11.78-i 0.33 $ & $0.72-i 0.01$ & $0.01 +i0.79$ & $ 0.01-i0.02$ & $ 0.71+i0.71$ & $3.98\times 10^{1}$ &  $--$\\
      (6) & $1p_{1/2}$ & $-19.65$  & $19.23-i 0.03 $ & $0.94-i 0.00$ & $0.00 +i0.99$ & $ 0.00$       & $-0.71+i0.70$ & $2.92\times 10^{2}$ &  $--$\\
      (7) & $1p_{3/2}$ & $-23.61$  & $23.07-i 0.01 $ & $1.03-i 0.00$ & $0.00+i1.08$  & $ 0.00$       & $ 0.71-i0.71$ & $8.29\times 10^{2}$ &  $--$\\
      (8) & $1s_{1/2}$ & $-34.78$  & $34.13-i 0.00$  & $1.26-i 0.00$ & $0.00+i1.30$  & $ 0.00$       & $-0.71-i0.71$ & $2.72\times 10^{3}$ &  $--$ \\
      \multicolumn{10}{c}{--Shape resonance--}\\
      (9) & $f_{5/2}$ & $--$  & $8.01-i1.81$ & $0.59-i0.07$   & $0.07+i 0.66$ & $0.38+i0.59$ & $ 0.17-i0.69$ & $1.53\times 10^{-1}$    &  $--$\\
      (10)& $g_{9/2}$ & $--$  & $10.23-i1.21$ & $0.67-i0.04$   & $0.04+i 0.74$ & $0.29+i0.78$ & $ 0.50+i0.25$ & $7.96\times 10^{-2}$  &  $--$
    \end{tabular}
  \end{ruledtabular}
\end{table*}

The calculation is performed in the following three steps:

Table \ref{table1} summarizes the baseline properties in the absence of the pairing correlation ($\langle \Delta \rangle = 0.0$ MeV) and without complex scaling ($\theta = 0.0$).
In this limit, the quasiparticle resonance energies $E_{nlj}$ are determined by the relation $E_{nlj} = |\epsilon_{nlj} - \lambda|$, with $\lambda = -1.0$ MeV. The
states are classified into three categories: particle-type, hole-type, and shape resonances. It is found that at $\theta = 0.0$, the normalization condition [Eq. (118)]
is strictly satisfied (resulting in $1.00$) only for the bound-state-like particle-type quasiparticles, whereas it remains undefined for other resonant states due to
their divergent asymptotic behavior at $r \to \infty$. 

In Table \ref{table2}, we introduce a finite pairing correlation ($\langle \Delta \rangle = 3.0$ MeV) while maintaining $\theta = 0.0$. This stage demonstrates the transformation
of the single-particle levels into quasiparticle resonances with complex energies $E_{nlj}$. Since complex scaling is not yet applied, all the resonant wave functions
exhibit divergent behavior, rendering the normalization via Eq.~(\ref{normal}) non-integrable for all states, as indicated by the symbol "--".

Table \ref{table3} provides the definitive verification of the complex-scaled Jost-HFB method using a scaling angle of $\theta = 0.3$. By comparing these results with Table \ref{table2},
we numerically confirm the invariance of the quasiparticle resonance energies $E_{nlj}$ with respect to $\theta$. Furthermore, the complex scaling regularizes the
divergent wave functions, rigorously restoring the normalization to exactly $1.00$ for all resonant states. These results numerically prove that the combined use of the
complex-scaled Jost function and the Autonne-Takagi factorization provides a mathematically sound and stable definition of the non-Hermitian HFB eigenstates.

\begin{table*}
  \caption{
    Same as Table II, but with the complex scaling angle $\theta = 0.3$. The invariance of $E_{nlj}$ relative to Table II confirms the $\theta$-independence
    of the resonance poles. In the last column, the complex scaling regularizes the divergent wave functions, restoring the normalization to exactly $1.00$
    for all states according to Eq. (\ref{normal}). This numerical achievement proves the validity of the
    dual basis defined via the Autonne-Takagi factorization and the mathematical consistency of the proposed normalization scheme for the Jost-HFB framework.  }
  \label{table3}
  \begin{ruledtabular}
    \begin{tabular}{cccccccccc}
      & & & \multicolumn{7}{c}{$\lambda=-1.0$ MeV, $\bra\Delta\ket=3.0$ MeV, $\theta=0.3$}\\
      \cline{4-10}
      No. & $nlj$ & $\epsilon_{nlj}$
      & $E_{nlj}$ & $k_1^{\theta}(E_{nlj})$ & $k_2^{\theta}(E_{nlj})$
      & \multicolumn{1}{c}{$Q^{\theta}_{11}$} & \multicolumn{1}{c}{$Q^{\theta}_{21}$} & $\sigma^{\theta}_{nlj}$ 
      & \makecell[l]{$e^{+2i\theta}$\\$\times\bra\widetilde{\varphi}_{nlj}^{\theta}|\varphi_{nlj}^{\theta}\ket$}
      \\
      & & [MeV] & [MeV] & [fm$^{-1}$] & [fm$^{-1}$] & \multicolumn{1}{c}{[1]} & \multicolumn{1}{c}{[1]} & [fm$^{-2}$] & [1] \\
      \colrule
      \multicolumn{10}{c}{--particle-type quasiparticle resonance--}\\
      (1) & $2p_{3/2}$ & $ -0.21$  & $ 1.35-i 0.17 $ & $0.14+i0.01$ & $-0.09+i0.33$ & $ 0.11+i0.13$ & $-0.39+i0.91$ & $2.50\times 10^{-1}$ & $1.00$ \\
      (2) & $1f_{7/2}$ & $ -0.93$  & $ 2.45-i 0.01 $ & $0.25+i0.08$ & $-0.12+i0.39$ & $ 0.00+i0.16$ & $ 0.69-i0.71$ & $2.17\times 10^{-2}$ & $1.00$ \\
      \multicolumn{10}{c}{--hole-type quasiparticle resonance--}\\
      (3) & $1d_{3/2}$ & $ -5.09$  & $ 4.94-i 0.25 $ & $0.42+i0.12$ & $-0.15+i0.52$ & $ 0.01+i0.08$ & $-0.76-i0.64$ & $1.64$               & $1.00$ \\
      (4) & $2s_{1/2}$ & $ -8.62$  & $ 8.22-i 0.48 $ & $0.57+i0.16$ & $-0.18+i0.64$ & $-0.01+i0.01$ & $-0.73-i0.68$ & $8.73\times 10^{1}$   & $1.00$ \\
      (5) & $1d_{5/2}$ & $-12.17$  & $11.78-i 0.33 $ & $0.69+i0.20$ & $-0.22+i0.75$ & $ 0.01-i0.02$ & $ 0.71+i0.71$ & $3.98\times 10^{1}$   & $1.00$ \\
      (6) & $1p_{1/2}$ & $-19.65$  & $19.23-i 0.03 $ & $0.90+i0.28$ & $-0.29+i0.94$ & $ 0.00$       & $-0.71+i0.70$ & $2.92\times 10^{2}$   & $1.00$ \\
      (7) & $1p_{3/2}$ & $-23.61$  & $23.07-i 0.01 $ & $0.99+i0.30$ & $-0.32+i1.03$ & $ 0.00$       & $ 0.71-i0.71$ & $8.29\times 10^{2}$   & $1.00$ \\
      (8) & $1s_{1/2}$ & $-34.78$  & $34.13-i 0.00$ & $1.21+i0.37$ & $-0.38+i1.24$ & $ 0.00$       & $-0.71-i0.71$ & $2.72\times 10^{3}$   & $1.00$\\
      \multicolumn{10}{c}{--Shape resonance--}\\
      (9) & $f_{5/2}$ & $--$  & $8.01-i1.81$  & $0.58+i0.10$   & $-0.13+i 0.65$ & $0.38+i0.59$ & $ 0.17-i0.69$ & $1.53\times 10^{-1}$    & $1.00$\\
      (10)& $g_{9/2}$ & $--$  & $10.23-i1.21$ & $0.65+i0.16$   & $-0.18+i 0.72$ & $0.29+i0.78$ & $ 0.50+i0.25$ & $7.96\times 10^{-2}$    & $1.00$
    \end{tabular}
  \end{ruledtabular}
\end{table*}

The numerical results presented in Tables \ref{table1}, \ref{table2}, and \ref{table3} reveal several crucial aspects of the proposed complex-scaled Jost-HFB framework.
A primary hallmark of our method is the numerical invariance of the resonance properties with respect to the complex scaling angle $\theta$.
By comparing Table \ref{table2} ($\theta = 0.0$) and Table \ref{table3} ($\theta = 0.3$), it is evident that not only the complex quasiparticle energies $E_{nlj}$ but
also the Takagi factors—the unitary matrix elements ($Q_{11}^\theta, Q_{21}^\theta$) and the singular values $\sigma_{nlj}^\theta$—remain strictly identical.
This numerical invariance is a direct manifestation of the analytical property of the Jost function as given in Eq. (\ref{invarianceJost}).
Since the Jost function is invariant under complex scaling when expressed as a function of the complex energy $E$, the $S$-matrix and its residues at the resonance poles
$\mathcal{M}_{nlj}^\theta$ [Eq. (\ref{Mdef})] must also be independent of $\theta$. 
Consequently, the factors obtained from the Autonne-Takagi factorization of these residues are naturally $\theta$-invariant.
This consistency numerically demonstrates that our framework provides a robust and unique description of the internal structure of quasiparticle resonances.

The introduction of the pairing correlation ($\langle \Delta \rangle = 3.0$ MeV) clearly illustrates the mechanism of ``particle-hole mixing'' within the resonant states.
In the absence of pairing (Table \ref{table1}), the particle-type states consist solely of the upper component ($Q_{21} = 0$), while the hole-type states consist only of the lower
component ($Q_{11} = 0$). Upon introducing the pairing interaction (Tables \ref{table2} and \ref{table3}), both $Q_{11}^\theta$ and $Q_{21}^\theta$ acquire finite values for all states.
This indicates that the traditional HFB concept of $u, v$ mixing coefficients is rigorously extended to non-Hermitian resonance states through the Autonne-Takagi factorization. 

Furthermore, the transition from bound single-particle levels to quasiparticle resonances is clearly captured. As shown in Table \ref{table1}, several single-particle levels
(particularly the hole-type states) are initially bound. However, the introduction of the pairing correlation in Tables \ref{table2} and \ref{table3} transforms these levels into
quasiparticle resonances with complex energies $E_{nlj}$. The finite imaginary parts of $E_{nlj}$ represent the decay widths induced by the pairing interaction, which
couples the bound states to the continuum. Notably, the decay widths for these pairing-induced resonances (especially for hole-type states) are significantly smaller
($\Gamma \approx 0.5$ MeV or less) compared to the typical shape resonances ($f_{5/2}, g_{9/2}$), suggesting that they exist as relatively long-lived, isolated
excitations. 

The singular values $\sigma_{nlj}^\theta$ provide a quantitative measure of the residue strength for each pole.
These values vary across several orders of magnitude, from $10^{-4}$ for threshold-neighboring states to $10^{3}$ for deeply bound hole-type states like $1s_{1/2}$.
This wide variation reflects the different physical roles each resonance pole plays in the spectral representation of the Green's function and the density of states. 

Finally, the success of the normalization procedure is verified in the last column of Table \ref{table3}.
The application of complex scaling ensures that the imaginary parts of the momenta satisfy $\text{Im}\,k_1^\theta > 0$ and $\text{Im}\,k_2^\theta > 0$ for all resonant states.
This mathematical condition regularizes the divergent asymptotic behavior of the resonant wave functions, allowing them to acquire $L^2$-like properties. As a result,
the normalization integral based on the dual basis [Eq.(\ref{normal})] is numerically restored to exactly $1.00$ for all states.
This achievement, combined with the $\theta$-invariance discussed above, provides definitive numerical proof of the mathematical consistency and practical stability of
our complex-scaled Jost-HFB framework.

\begin{figure*}[htbp]
  \includegraphics[width=\linewidth]{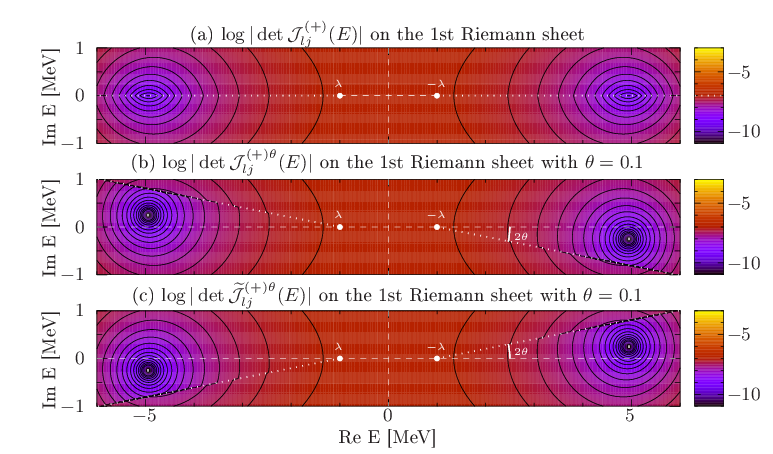}
  \caption{(Color online)
    Contour plots of $\log |\det \bvec{\mathcal{J}}_{lj}^{(+)}(E)|$ for the $1d_{3/2}$ state on the first Riemann sheet of the complex energy plane,
    calculated with the pairing gap $\langle \Delta \rangle = 3.0$ MeV and chemical potential $\lambda = -1.0$ MeV. Panel (a) shows the result without complex scaling
    ($\theta = 0.0$), while panels (b) and (c) show the results with $\theta = 0.1$ for the original and dual bases, respectively. The dashed lines starting from $E =
    -\lambda$ represent the branch cuts, which are rotated by $-2\theta$ in (b) and $+2\theta$ in (c). The smooth behavior of the contour lines approaching the branch cuts
    provides the basis for the analytical connection to the second Riemann sheet shown in Fig.~\ref{fig_mod2-detJ1}.  }
  \label{fig_mod1-detJ1}
\end{figure*}

\begin{figure*}[htbp]
  \includegraphics[width=\linewidth]{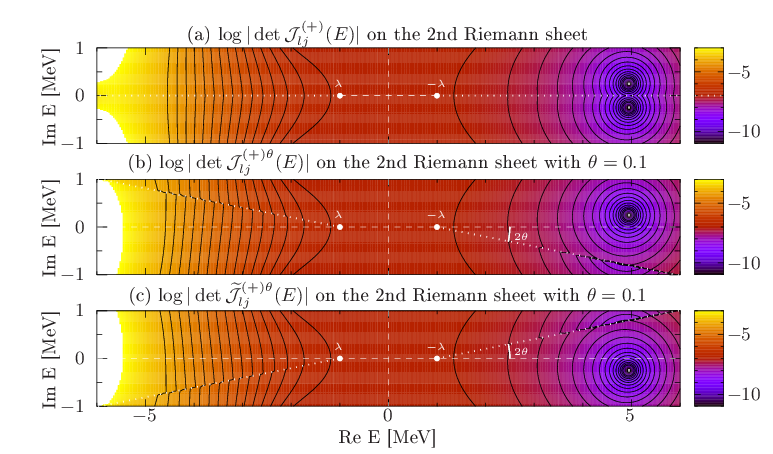}
  \caption{(Color online)
    Same as Fig.~\ref{fig_mod1-detJ1}, but on the second Riemann sheet. The resonance pole of the $1d_{3/2}$ state is clearly identified as a distinct zero point
    (dark region) at $E \approx 4.94 - i0.25$ MeV, which is consistent with the values listed in Tables \ref{table2} and \ref{table3}. Comparison with Fig.~\ref{fig_mod1-detJ1}
    through the contour consistency across the branch cuts demonstrates the seamless analytical connection between the two Riemann sheets.
    The position of the resonance pole remains strictly identical across panels (a), (b), and (c), numerically confirming the $\theta$-invariance of the resonance poles as
    analytically described by Eq.~(\ref{invarianceJost}).}
  \label{fig_mod2-detJ1}
\end{figure*}

\subsection{Analytical structure and Riemann surface of the Jost function}

To provide deeper insight into the analytical properties of the Jost function and to visually confirm the consistency of our framework, we present the contour plots of $\log |\det \bvec{\mathcal{J}}_{lj}^{(+)}(E)|$ for the $1d_{3/2}$ quasiparticle state in Figs. \ref{fig_mod1-detJ1} and \ref{fig_mod2-detJ1}. These figures illustrate the behavior of the Jost function on the first and second Riemann sheets, respectively.

As shown in Fig. \ref{fig_mod1-detJ1}, the resonance pole is not directly visible on the first Riemann sheet, as it is located on the unphysical sheet hidden behind the branch cut. However, on the second Riemann sheet (Fig. \ref{fig_mod2-detJ1}), the resonance pole is clearly identified as a distinct zero point (the dark region) at $E = 4.94 - i0.25$ MeV, which is in excellent agreement with the values listed in Tables \ref{table2} and \ref{table3}. Notably, the consistency of the contour patterns approaching the branch cuts (the dashed lines) in both figures demonstrates the seamless analytical connection between the two Riemann sheets. This continuity proves that the Jost function is correctly defined as a single analytical entity across the different sheets of the complex energy plane.

The comparison between panels (a) and (b) in both figures provide a geometric verification of the $\theta$-invariance of the resonance poles. While the complex scaling with $\theta = 0.1$ rotates the branch cuts by an angle of $-2\theta$ (clockwise) in panel (b), the complex energy of the resonance pole remains strictly identical to that in the $\theta = 0.0$ case (panel a). This numerical result is a direct manifestation of the analytical property expressed in Eq. (\ref{invarianceJost}). Furthermore, the rotation of the branch cut exposes the complex plane structure around the resonance pole that was previously "masked" by the branch cut on the real axis at $\theta = 0.0$. This exposure clearly explains the mechanism by which the complex scaling method enhances the numerical stability and convergence of the pole-searching algorithms.

Furthermore, the symmetry of the contour patterns across the Re $E$ axis (relative to the midpoint between $\lambda$ and $-\lambda$) provides visual evidence of the energy-inversion property described by Eq. (\ref{detjost-neg}). The presence of the resonance pole in the positive energy region is mirrored by a corresponding symmetric structure in the negative energy region. This reflects the existence of a symmetric pole at $E = -E_n$, which would be fully uncovered by crossing the branch cut associated with the momentum $k_2$ in the negative energy region ($E < \lambda$). The global consistency of these contours confirms that the complex-scaled Jost-HFB framework maintains its analytical integrity across the multiple Riemann sheets associated with both $k_1^\theta$ and $k_2^\theta$.

Finally, panel (c) in Figs. \ref{fig_mod1-detJ1} and \ref{fig_mod2-detJ1} demonstrates the properties of the dual basis. In accordance with the theoretical derivations in Sec.\ref{original_vs_dual}, the branch cut in the dual space is rotated in the opposite direction ($+2\theta$, counter-clockwise). Despite this reverse rotation, the resonance pole position remains invariant and perfectly coincides with those in the original space. This visual confirmation reinforces the mathematical rigor of the dual space definition and the validity of the Autonne-Takagi factorization as a basis for the rigorous normalization of non-Hermitian HFB states.

\begin{figure}[htbp]
  \includegraphics[width=\linewidth]{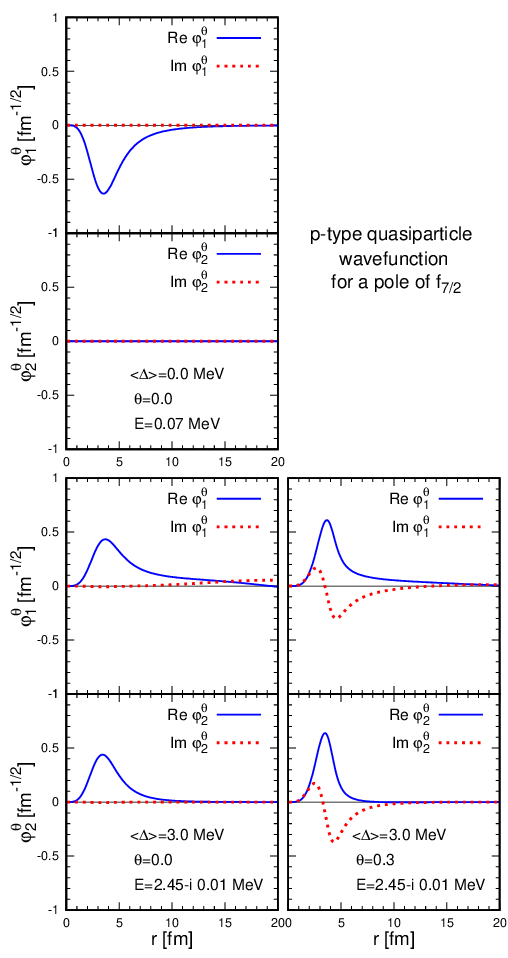}
  \caption{(Color online)
    Radial wave functions $\varphi_{1}^\theta(r)$ and $\varphi_{2}^\theta(r)$ for the p-type quasiparticle resonance state of $1f_{7/2}$ (pole No. 2 in
    Table \ref{table2} and \ref{table3}). The top panel shows the wave functions in the single-particle limit ($\langle \Delta \rangle = 0.0$ MeV, $\theta = 0.0$). The bottom-left and
    bottom-right panels show the results with the pairing correlation ($\langle \Delta \rangle = 3.0$ MeV) for the unscaled ($\theta = 0.0$) and complex-scaled
    ($\theta=0.3$) cases, respectively. The complex scaling regularizes the divergent behavior of the resonance wave function at large distances, ensuring its
    convergence as $r \to\infty$.  }
  \label{fig-wf_f7-2}
\end{figure}

\begin{figure}[htbp]
  \includegraphics[width=\linewidth]{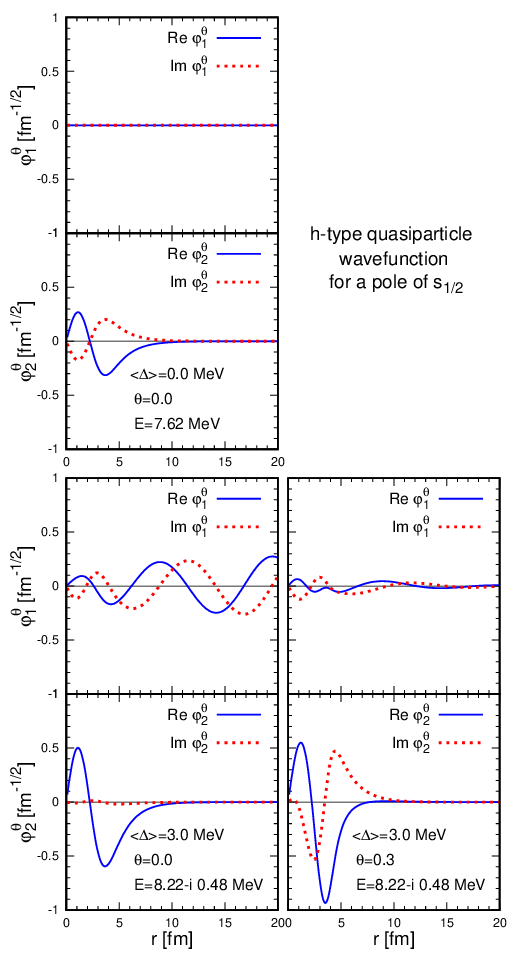}
  \caption{(Color online)
    Radial wave functions $\varphi_{1}^\theta(r)$ and $\varphi_{2}^\theta(r)$ for the h-type quasiparticle resonance state of $2s_{1/2}$ (pole No. 4 in
    Table \ref{table2} and \ref{table3}). The parameters and panel layout are the same as in Fig.\ref{fig-wf_f7-2}.
    For the h-type resonance, the lower component $\varphi_{2}^\theta$ is dominant, and its
    non-integrable behavior at $\theta = 0.0$ is effectively regularized to a square-integrable form by the complex scaling at $\theta = 0.3$.  }
  \label{fig-wf_s1-2}
\end{figure}

\begin{figure}[htbp]
  \includegraphics[width=\linewidth]{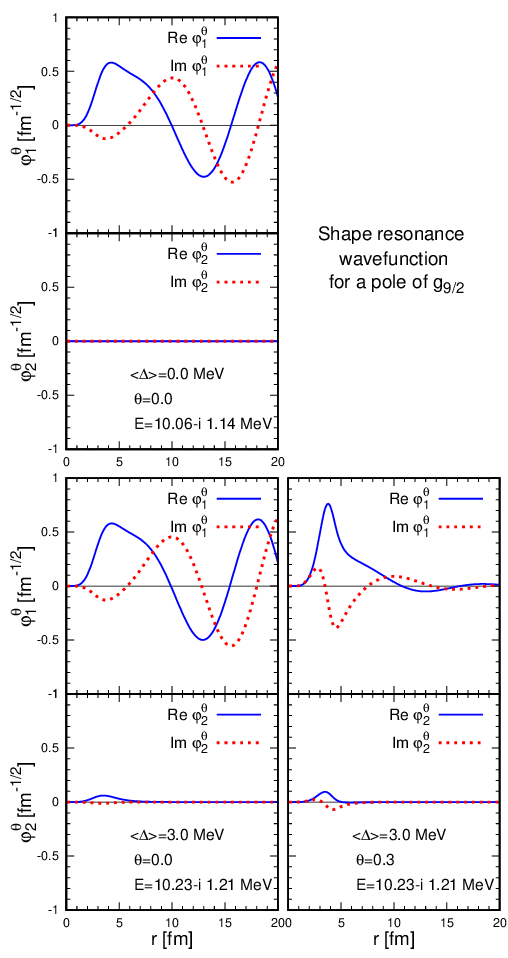}
  \caption{(Color online)
    Radial wave functions $\varphi_{1}^\theta(r)$ and $\varphi_{2}^\theta(r)$ for the shape resonance state of $1g_{9/2}$ (pole No. 10 in Table \ref{table2} and \ref{table3}).
    The parameters and panel layout follow those in Fig.\ref{fig-wf_f7-2}. The divergent tail of the shape resonance is clearly damped by the complex scaling ($\theta = 0.3$),
    confirming the mathematical consistency of the proposed normalization scheme for all types of HFB resonances.  }
  \label{fig-wf_g9-2}
\end{figure}

\subsection{Regularization and mixing properties of the resonant wave functions}

To provide a visual confirmation of the proposed framework, we examine the radial wave functions of the HFB resonant states. Figs.~\ref{fig-wf_f7-2}, ~\ref{fig-wf_s1-2},
and \ref{fig-wf_g9-2} display the two-component spinor wave functions $\varphi_{nlj}^\theta(r) = (\varphi_{nlj,1}^\theta(r), \varphi_{nlj,2}^\theta(r))^T$ defined in Eq.~(\ref{eigenwf}),
corresponding to the p-type, h-type, and shape resonance states, respectively. In these plots, the upper component $\varphi_{nlj,1}^\theta$ represents the particle-like contribution, while
the lower component $\varphi_{nlj,2}^\theta$ represents the hole-like contribution. These eigenfunctions are determined through the Autonne-Takagi factorization of the
residue matrix $\mathcal{M}_{nlj}^\theta$ [Eq. (\ref{takagidecom})] and are plotted with their real (solid lines) and imaginary (dotted lines) parts.

The effect of the pairing correlation and the resulting mixing properties are clearly visible when comparing the top panels ($\langle\Delta\rangle = 0.0$ MeV) with the
bottom panels ($\langle\Delta\rangle = 3.0$ MeV). In the absence of pairing, the wave functions are purely particle-like or hole-like. However, the introduction of
pairing leads to a significant mixing of the upper and lower components. This mixing is quantitatively characterized by the coefficients $Q_{11}$ and $Q_{21}$ listed in
Tables \ref{table1}–\ref{table3}. For instance, in the p-type resonance (Fig.~\ref{fig-wf_f7-2}), the dominance of $\varphi_1$ over $\varphi_2$ corresponds to the
large magnitude of $Q_{11}$ relative to $Q_{21}$ in Table \ref{table3}. Conversely, in the h-type resonance (Fig.~\ref{fig-wf_s1-2}), the lower component $\varphi_2$
exhibits a significantly larger amplitude, consistent with the dominance of $Q_{21}$. The fact that both components contribute to the total density demonstrates that
the Autonne-Takagi factorization effectively extracts the quasiparticle mixing amplitudes for non-Hermitian resonant states.

Furthermore, the convergence of these wave functions is directly linked to the complex momenta $k_1^\theta$ and $k_2^\theta$ shown in Tables \ref{table2} and \ref{table3}.
Asymptotically, the resonant solutions follow the behavior $\varphi_{nlj}^\theta(r) \sim e^{i k_{1,2}^\theta r}$. For the wave function to be square-integrable ($L^2$),
the imaginary part of the momentum must be positive ($\text{Im } k_{1,2}^\theta > 0$). In the unscaled case ($\theta = 0.0$, Table \ref{table2}), the resonance poles
possess negative imaginary momenta, leading to the exponentially growing tails observed in the bottom-left panels of Figs. ~\ref{fig-wf_f7-2}–~\ref{fig-wf_g9-2}.
This divergence is the reason why the normalization integral remains undefined in Table \ref{table2}.
In contrast, the application of complex scaling rotates the momenta into the upper-half of the complex $k$-plane. As shown in Table \ref{table3}($\theta = 0.3$),
$\text{Im } k_1^\theta$ and $\text{Im } k_2^\theta$ become positive for all resonant states, resulting in the rapid damping of the wave function tails as $r \to \infty$
(bottom-right panels).

A paramount feature of the present theory is that the normalization of the resonant wave functions is achieved automatically rather than artificially. In conventional
complex scaling applications, the wave functions are typically normalized post-hoc by numerically evaluating $\int \phi^2 dr$ and rescaling the amplitude. In our
approach, however, the normalization is intrinsically embedded in the mathematical structure of the residue matrix $\bvec{\mathcal{M}}_{nlj}^\theta$ and its Autonne-Takagi
factorization. The eigenfunctions $\varphi_{nlj}^\theta$ are uniquely determined with the correct absolute scale from the outset. The fact that the normalization
integrals in the last column of Table \ref{table3} yield exactly $1.00$ without any artificial adjustments provides a rigorous numerical proof of the self-normalizing nature and
the mathematical consistency of the complex-scaled Jost-HFB framework. 

\begin{figure*}[htbp]
  \includegraphics[width=\linewidth]{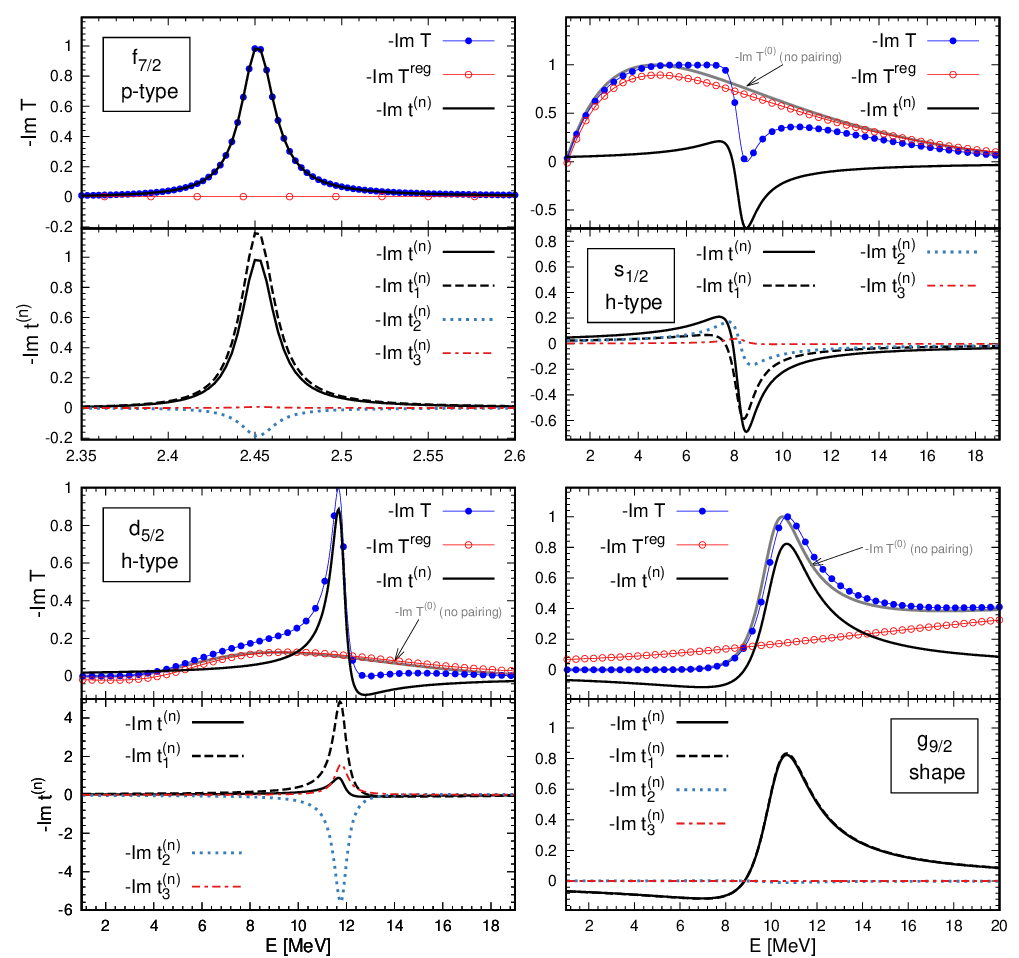}
  \caption{(Color online)
    Decomposition of the T-matrix into the pole contribution and the background term for the $f_{7/2}$ (p-type), $s_{1/2}$ (h-type), $d_{5/2}$
    (h-type), and $g_{9/2}$ (shape) quasiparticle resonances. In each panel, the top sub-plot shows the total $-\text{Im } T$ (blue dots), the regular background term
    $-\text{Im } T^{\text{reg}}$ (red circles), and the total pole contribution $-\text{Im } t^{(n)}$ (black solid line) based on the Mittag-Leffler expansion [Eq. (\ref{MLtheorem})].
    The bottom sub-plot shows the decomposition of the pole contribution into three terms defined in Eq. (\ref{resTmat}): the mean-field scattering term $-\text{Im } t^{(n)}_1$
    (black solid line), the interference term $-\text{Im } t^{(n)}_2$ (blue dotted line), and the pairing scattering term $-\text{Im } t^{(n)}_3$ (red dashed line).  }
  \label{fig-Tmat}
\end{figure*}

\begin{table*}
  \caption{
    Decomposition of the T-matrix residues into the mean-field scattering, interference, and pairing scattering components according to Eq. (\ref{resTmat}). The magnitudes
  and the phases (where $\delta_n^{(1)}$ and $\delta_n^{(2)}$ represent the arguments of $\gamma_n^{(1)}$ and $\gamma_n^{(2)}$, respectively) are listed for each
  resonance state. The rightmost two columns compare the residues calculated via numerical contour integration of the T-matrix [Eq. (\ref{resT})] and the analytical expression
  based on the resonant wave function integral [Eq. (\ref{resT2})]. The exact numerical agreement between these two columns, which holds independently of the scaling angle
  $\theta$, confirms the mathematical consistency of the residue representation and the pole-decomposition of the T-matrix within the HFB framework.  }
  \label{table4}
  \begin{ruledtabular}
    \begin{tabular}{ccccccccccc}
      & & &\multicolumn{8}{c}{$\lambda=-1.0$ MeV, $\bra\Delta\ket=3.0$ MeV, $\theta=0.0$ and $0.3$}\\
      \cline{3-11}
      & & & & & &&&&\multicolumn{2}{c}{$\gamma_n^2$=Res$\left[T_{lj}^{\theta},E_{nlj}\right]$} \\
      \cline{10-11}
      No. & $nlj$ 
      & $E_{nlj}$ & $|\gamma_n^{(1)}|^2$ & $2\delta_n^{(1)}$ & $2|\gamma_n^{(1)}||\gamma_n^{(2)}|$ & $\delta_n^{(1)}+\delta_n^{(2)}$ 
      & $|\gamma_n^{(2)}|^2$ & $2\delta_n^{(2)}$ & Eq.(\ref{resT}) & Eq.(\ref{resT2}) \\
      & & [MeV] & [MeV] & [rad] & [MeV] & [rad] & [MeV] & [rad] &  [$\times 10^{-2}$MeV] & [$\times 10^{-2}$MeV] \\
      \colrule
      \multicolumn{11}{c}{--particle-type quasiparticle resonance--}\\
      (1) & $2p_{3/2}$ & $ 1.35-i 0.17 $ & $0.17$  & $-0.43\pi$ & $0.031$ & $0.59\pi$ & $1.4\times 10^{-3}$ & $-0.39\pi$ & $3.05-i 14.2$              & $3.05-i 14.2$ \\
      (2) & $1f_{7/2}$ & $ 2.45-i 0.01 $ & $0.01$  & $-0.02\pi$ & $0.002$ & $0.98\pi$ & $0.1\times 10^{-3}$ & $-0.01\pi$ & $1.16-i 5.95\times 10^{-2}$ & $1.16-i 5.95\times 10^{-2}$ \\
      \multicolumn{11}{c}{--hole-type quasiparticle resonance--}\\
      (3) & $1d_{3/2}$ & $ 4.94-i 0.25 $ & $0.60$ & $-0.09\pi$ & $0.42$ & $ 0.92\pi$ & $0.08$ & $-0.06\pi$ & $ 23.59  -i 7.91$  & $ 23.59  -i 7.91$  \\ 
      (4) & $2s_{1/2}$ & $ 8.22-i 0.48 $ & $0.31$ & $ 0.79\pi$ & $0.16$ & $ 0.49\pi$ & $0.02$ & $ 0.20\pi$ & $-22.86  +i 36.36$ & $-22.86  +i 36.36$ \\ 
      (5) & $1d_{5/2}$ & $11.78-i 0.33 $ & $1.64$ & $ 0.09\pi$ & $1.86$ & $-0.96\pi$ & $0.53$ & $ 0.00$    & $ 26.11  +i 19.65$ & $ 26.11  +i 19.65$ \\ 
      (6) & $1p_{1/2}$ & $19.23-i 0.03 $ & $2.42$ & $ 0.00$    & $4.28$ & $-1.00\pi$ & $1.89$ & $ 0.00$    & $ 3.26   -i 0.23$  & $ 3.26   -i 0.23$  \\ 
      (7) & $1p_{3/2}$ & $23.07-i 0.01 $ & $1.94$ & $ 0.00$    & $3.63$ & $-1.00\pi$ & $1.70$ & $ 0.00$    & $ 0.75   +i 0.10$  & $ 0.75   +i 0.10$  \\ 
      (8) & $1s_{1/2}$ & $34.13-i 0.00$ & $0.31$ & $ 0.02\pi$ & $0.67$ & $-0.99\pi$ & $0.37$ & $ 0.00$    & $ 0.23   -i 0.16$  & $ 0.23   -i 0.16$  \\ 
      \multicolumn{11}{c}{--Shape resonance--}\\
      (9) & $f_{5/2}$ & $8.01-i1.81$  & $1.62$ & $-0.36\pi$ & $0.06$ & $0.63\pi$ & $0.00$ & $-0.38\pi$ & $66.07-i 141.2$  & $66.07-i 141.2$ \\
      (10)& $g_{9/2}$ & $10.23-i1.21$ & $1.15$ & $-0.23\pi$ & $0.01$ & $0.76\pi$ & $0.00$ & $-0.26\pi$ & $85.88-i 74.04$  & $85.88-i 74.04$ \\
    \end{tabular}
  \end{ruledtabular}
\end{table*}

\subsection{Numerical Analysis of the $T$-matrix Residues and Scattering Profiles}
Table~\ref{table4} summarizes the numerical decomposition of the $T$-matrix residues $\gamma_n^2$ for each quasiparticle resonance, based on the analytical
representation given in Eq.~(\ref{resTmat}). The table lists the magnitudes of the mean-field scattering component $|\gamma_n^{(1)}|^2$, the pairing scattering
component $|\gamma_n^{(2)}|^2$, and the interference term $2|\gamma_n^{(1)}||\gamma_n^{(2)}|$, along with their respective phases ($2\delta_n^{(1)}$,
$\delta_n^{(1)}+\delta_n^{(2)}$ and $2\delta_n^{(2)}$). To verify the
mathematical consistency of our framework, the rightmost two columns compare the residues calculated via numerical contour integration of the $T$-matrix
[Eq.~(\ref{resT})] and the analytical expression using the complex-scaled wave functions obtained from the Autonne-Takagi factorization [Eq.~(\ref{resT2})].
Complementing these values, Fig.~\ref{fig-Tmat} displays the energy dependence of the $T$-matrix $-\text{Im } T$ for representative states ($1f_{7/2}, 2s_{1/2},
1d_{5/2}, 1g_{9/2}$), illustrating the decomposition into the pole contribution and the regular background term based on the Mittag-Leffler expansion
[Eq.~(\ref{MLtheorem})].

A primary hallmark of the present results is the exact numerical agreement between the two residue calculation methods shown in Table~\ref{table4}. This consistency,
which holds identically for both $\theta = 0.0$ and $0.3$, provides a rigorous numerical proof that the $S$-matrix and $T$-matrix are invariant under complex scaling
when expressed as a function of the complex energy $E$, as analytically predicted in Eq.~(\ref{invarianceJost}). This result confirms that the complex-scaled Jost-HFB
framework correctly preserves the analytical structure of the scattering matrix across the Riemann sheets, ensuring that the non-Hermitian resonant states are uniquely
and robustly defined as physical eigenstates.

The physical nature of these resonances is clearly revealed through the decomposition of the pole contribution in Fig.~\ref{fig-Tmat}. For the $2s_{1/2}$ state (pole
No.~4), Table~\ref{table4} shows a dominant mean-field scattering term ($|\gamma_n^{(1)}|^2 = 0.31$ MeV) compared to a significantly smaller pairing term
($|\gamma_n^{(2)}|^2 = 0.02$ MeV). Because $s$-wave particles lack a centrifugal barrier, they experience strong background scattering from the mean-field potential.
Notably, the phase $2\delta_n^{(1)}$ is approximately $0.79\pi$, which is close to $\pi$ (anti-phase). This phase relation leads to a strong destructive interference
between the background scattering and the hole-type resonance, resulting in the characteristic ``Fano dip'' observed in Fig.~\ref{fig-Tmat} for the $s_{1/2}$ state.

In contrast, for the $1d_{5/2}$ state (pole No.~5), the phase $2\delta_n^{(1)}$ is as small as $0.09\pi$, indicating a nearly in-phase (constructive) interference.
Combined with the presence of the centrifugal barrier ($l=2$) that tends to confine the resonance, the $d_{5/2}$ state exhibits an asymmetric peak profile, often
referred to as a Fano resonance. These differences between $s_{1/2}$ and $d_{5/2}$ clearly demonstrate that the angular momentum $l$ governs the interference phase and
the relative strength of the background, thereby diversifying the scattering observables.

Furthermore, the results for the $1f_{7/2}$ (particle-type) and $1g_{9/2}$ (shape resonance) states provide a clear distinction from the hole-type excitations. As shown
in Table~\ref{table4}, the $1f_{7/2}$ state is characterized by an extremely small pairing scattering term ($|\gamma_n^{(2)}|^2 = 0.1 \times 10^{-3}$ MeV).
Consequently, its $T$-matrix profile in Fig.~\ref{fig-Tmat} shows a sharp, symmetric Lorentzian-like peak, where the interference effect is almost negligible. The
$1g_{9/2}$ state, on the other hand, is a pure shape resonance originating from the mean-field barrier, with the pairing scattering term being exactly zero
($|\gamma_n^{(2)}|^2 = 0.00$). Its $T$-matrix exhibits a broad peak structure without any dip, reflecting a pure potential scattering process.

Finally, while the asymmetric $T$-matrix profiles and dips observed in our analysis are naturally explained by the interference terms within the Mittag-Leffler
expansion, it should be noted that a rigorous mathematical mapping between our complex residue components ($\gamma_n^{(1)}, \gamma_n^{(2)}$) and the phenomenological
Fano resonance formula (specifically the asymmetry parameter $q$) remains to be established. While the qualitative behavior strongly suggests a Fano-type mechanism, a
formal derivation that links the Jost-HFB residues directly to the standard Fano parameters is a significant subject for future research. The current work provides the
foundational numerical evidence for such interference in open HFB systems, opening a path toward a unified understanding of Fano-type phenomena in quasiparticle
scattering.

\section{Summary and outlook}
In this study, we have presented a theoretical framework to describe quasiparticle resonance states within the Hartree-Fock-Bogoliubov (HFB)
theory by combining the complex-scaled Jost function method with the Autonne-Takagi factorization. This approach provides a consistent formulation
for the completeness relation and the normalization of the resonant wave functions in the coupled-channel HFB equations.

The completeness relation for the HFB system was derived from the analytical properties of the HFB Green's function through contour integration in
the complex energy plane. We demonstrated that the introduction of the complex scaling method (CSM) is essential for explicitly separating the
resonance pole contributions from the continuum background, ensuring that the resonant states are properly included as discrete components in the
completeness relation.

To define and normalize the resonant wave functions (Gamow states) at the S-matrix poles, we utilized the complex symmetry of the flux-adjusted
S-matrix. Since the residue matrix of the S-matrix becomes rank-1 at the pole energy, the application of the Autonne-Takagi factorization uniquely
determines the resonant wave function and its normalization factor. This scheme allows for the construction of eigenfunctions with a correct
absolute scale without relying on artificial adjustments or phenomenological basis sets.

Numerical analysis confirmed the mathematical consistency and stability of the present formalism. We verified that the resonance energies and the
Takagi-normalized wave functions remain invariant under the rotation of the complex scaling angle $\theta$. Furthermore, the T-matrix residues
calculated via the Mittag-Leffler expansion were found to be in exact numerical agreement with those obtained from the microscopic integrals of
the defined Gamow states.

The physical characteristics of the quasiparticle resonances were investigated through the scattering profiles of several orbitals ($s_{1/2},
d_{5/2}, f_{7/2}, g_{9/2}$). The analysis showed that the diverse resonance structures, such as the characteristic Fano dips in the $s_{1/2}$
channel and the asymmetric peaks in the $d_{5/2}$ channel, originate from the interference between the discrete resonance poles and the background
continuum. These results indicate that hole-type resonances in the HFB framework can be understood as a manifestation of the Fano process.

The theoretical framework established in this work, particularly the normalization method using the Autonne-Takagi factorization, can be naturally
extended to the Jost-RPA framework. This extension will enable the unique definition and normalization of transition densities for collective
excitations in the continuum, providing a microscopic foundation for evaluating the collectivity of various collective modes---including those
expected in nuclei far from stability---within open quantum many-body systems.

\section*{Acknowledgment}
The author is grateful to Shoya Ogawa (Kyushu University) for fruitful discussions on the
foundations and conceptual understanding of the complex scaling method. 

%


\begin{thebibliography}{00}
\bibitem{Ring1980} P. Ring and P. Schuck, \textit{The Nuclear Many-Body Problem} (Springer-Verlag, New York, 1980).
\bibitem{Bulgac1980} A. Bulgac, arXiv:nucl-th/9907088 (1980); A. Bulgac, IPNE-FT-194-1980 (1980).
\bibitem{Dobaczewski1984} J. Dobaczewski, H. Flocard, and J. Treiner, Nucl. Phys. A \textbf{422}, 103 (1984).
\bibitem{Belyaev1987} S. T. Belyaev, A. V. Smirnov, S. V. Tolokonnikov, and S. A. Fayans, Sov. J. Nucl. Phys. \textbf{45}, 783 (1987).
\bibitem{Matsuo2001} M. Matsuo, Nucl. Phys. A \textbf{696}, 371 (2001).
\bibitem{ABC1} J. Aguilar and J. M. Combes, Commun. Math. Phys. \textbf{22}, 269 (1971).
\bibitem{ABC2} E. Balslev and J. M. Combes, Commun. Math. Phys. \textbf{22}, 280 (1971).
\bibitem{Simon1973} B. Simon, Ann. Math. \textbf{97}, 247 (1973).
\bibitem{Moiseyev1998} N. Moiseyev, Phys. Rep. \textbf{302}, 212 (1998).
\bibitem{Aoyama2006} S. Aoyama, T. Myo, K. Kat\={o}, and K. Ikeda, Prog. Theor. Phys. \textbf{116}, 1 (2006).
\bibitem{Michel2002} N. Michel, W. Nazarewicz, M. P{\l}oszajczak, and K. Bennaceur, Phys. Rev. Lett. \textbf{89}, 042502 (2002).
\bibitem{Michel2009} N. Michel, W. Nazarewicz, M. P{\l}oszajczak, and T. Vertse, J. Phys. G: Nucl. Part. Phys. \textbf{36}, 013101 (2009).
\bibitem{Berggren1968} T. Berggren, Nucl. Phys. A \textbf{109}, 265 (1968).
\bibitem{Mizuyama2019} K. Mizuyama, N. N. Le, T. D. Thuy, and T. V. N. Hao, Phys. Rev. C \textbf{99}, 054607 (2019).
\bibitem{Mizuyama2024} K. Mizuyama, T. D. Thuy, and T. V. N. Hao, Phys. Rev. C \textbf{109}, 054304 (2024).
\bibitem{Mizuyama2025} K. Mizuyama and T. D. Thuy, Phys. Rev. C \textbf{111}, 054323 (2025).
\bibitem{Autonne1915} L. Autonne, Ann. Univ. Lyon, Nouv. S\'{e}r. I, Fasc. \textbf{38}, 1 (1915).
\bibitem{Takagi1925} T. Takagi, Jpn. J. Math. \textbf{1}, 83 (1925).
\bibitem{Rakityansky} Sergei A. Rakityansky, {\it Jost functions in Quantum Mechanics},
  {\it Springer}.
\end{thebibliography}
\end{document}